\documentclass[11pt,letterpaper]{article}
																																  \usepackage[portrait,margin=2.5cm]{geometry}

																																  \usepackage{setspace}
																																  \usepackage{titlesec}
																																  \usepackage[titles]{tocloft}
																																  \usepackage{epsfig}
																																  \usepackage{epstopdf}
																																  \usepackage{multirow}
																																  \usepackage[font=small,labelfont=bf]{caption}
																																  \usepackage{enumerate}
																																  \usepackage{dsfont}
																																  \usepackage{tikz}
																																  \usepackage{pgfplots}
																																  \usepackage{amssymb}
																																  \usepackage{subcaption}
																																  \usepackage[ruled,vlined]{algorithm2e}
																																  \usepackage{upgreek}
																																  \usepackage{afterpage}

																																  \usepackage[round]{natbib} 

																																  \usepackage[font=small,skip=-4pt]{subcaption}
																																  \usepackage{color}

																																  \usepackage{amsthm}
																																  \usepackage{amsmath}
																																  \usepackage{amssymb}

																																  \usepackage{xspace}

																																  \usepackage[hidelinks,linktocpage=true]{hyperref} 

\newcounter{Appendix}

\renewcommand{\cite}{\citep}

\theoremstyle{plain}

\newtheoremstyle{mydef}
        {3pt}
        {3pt}
        {\normalfont}
        {}
        {\bfseries}
        {.}
        {.5em}
        {\thmname{#1} \thmnumber{#2}\thmnote{#3}}

\newcommand\rightparend[1]{{%
      \unskip\nobreak\hfil\penalty50
      \hskip2em\hbox{}\nobreak\hfil\textbf{#1}%
      \parfillskip=0pt \finalhyphendemerits=0 \par}}

\newtheorem{xxexample}{Example}[section]

																																  \usepackage{epsfig,amssymb,amsmath, multirow, url, amsfonts, tcolorbox,booktabs, enumitem}

																																  \usepackage{amsthm}																									
																																  \usepackage[utf8]{inputenc}
																																  \usepackage{bbm}
																																  \usepackage{dsfont}

																																  \usepackage{hyperref}																								
						  \usepackage{graphicx}
																																  \usepackage{upref,setspace}
																																  \usepackage{caption}
																																  \usepackage{enumerate}

																																  \numberwithin{equation}{section}
																																  \numberwithin{figure}{section}
																																  \numberwithin{table}{section}

																																  \newtheorem{thm}{Theorem}[section]

																																  \newtheorem{prop}[thm]{Proposition}
																																  \newtheorem{defn}[thm]{Definition}

																																  \newtheorem*{ass*}{Assumption}
																																  
																																  \newtheorem{ex}[thm]{Example}

																																  \theoremstyle{definition}

																																  \renewcommand{\leq}{\leqslant}
																																  \renewcommand{\geq}{\geqslant}

																																  \newcommand{\wt}{\widetilde}
																																  \newcommand{\wh}{\widehat}

																																  \newcommand{\set}[1]{\left\{#1\right\}}

																																  \let\ga=\alpha \let\gb=\beta  \let\gd=\delta \let\gee=\epsilon
																																      \let\gk=\kappa \let\gl=\lambda \let\gm=\mu         \let\gr=\rho \let\gs=\sigma \let\gt=\tau 
																																    \let\gz=\zeta
																																   \let\gD=\Delta

																																  \newcommand{\cB}{\mathcal{B}}\newcommand{\cC}{\mathcal{C}}
																																  
																																  \newcommand{\cI}{\mathcal{I}}
																																  
																																  \newcommand{\cN}{\mathcal{N}}
																																  \newcommand{\cP}{\mathcal{P}}
																																  \newcommand{\cU}{\mathcal{U}}
																																  \newcommand{\cX}{\mathcal{X}}
																																  
																																  \newcommand{\vp}{\mathbf{p}}

																																  \newcommand{\mv}[1]{\boldsymbol{#1}}\newcommand{\mvzero}{\boldsymbol{0}}
																																  \newcommand{\mvone}{\boldsymbol{1}}

																																  \newcommand{\mvA}{\boldsymbol{A}}\newcommand{\mvB}{\boldsymbol{B}}\newcommand{\mvC}{\boldsymbol{C}}
																																  \newcommand{\mvD}{\boldsymbol{D}}\newcommand{\mvE}{\boldsymbol{E}}\newcommand{\mvF}{\boldsymbol{F}}
																																  \newcommand{\mvI}{\boldsymbol{I}}
																																  \newcommand{\mvJ}{\boldsymbol{J}}
																																  
																																  \newcommand{\mvP}{\boldsymbol{P}}\newcommand{\mvQ}{\boldsymbol{Q}}
																																  
																																  \newcommand{\mvV}{\boldsymbol{V}}\newcommand{\mvX}{\boldsymbol{X}}
																																  \newcommand{\mvY}{\boldsymbol{Y}}\newcommand{\mvZ}{\boldsymbol{Z}}\newcommand{\mva}{\boldsymbol{a}}
																																  \newcommand{\mvb}{\boldsymbol{b}}
																																  \newcommand{\mvf}{\boldsymbol{f}}
																																  
																																  \newcommand{\mvm}{\boldsymbol{m}}

																																  \newcommand{\mvx}{\boldsymbol{x}}\newcommand{\mvy}{\boldsymbol{y}}

																																  \newcommand{\mvgd}{\boldsymbol{\delta}}
																																  \newcommand{\mvgee}{\boldsymbol{\epsilon}}
																																  \newcommand{\mvgF}{\boldsymbol{\Phi}}
																																  \newcommand{\mvgh}{\boldsymbol{\eta}}
																																  \newcommand{\mvgl}{\boldsymbol{\lambda}}\newcommand{\mvgL}{\boldsymbol{\Lambda}}\newcommand{\mvgm}{\boldsymbol{\mu}}
																																  
																																  \newcommand{\mvgP}{\boldsymbol{\Pi}}
																																  \newcommand{\mvgS}{\boldsymbol{\Sigma}}
																																  
																																  \newcommand{\mvgz}{\boldsymbol{\zeta}}

																																  \newcommand{\mvlambda}{\boldsymbol{\lambda}}\newcommand{\mvLambda}{\boldsymbol{\Lambda}}\newcommand{\mvmu}{\boldsymbol{\mu}}

																																  \newcommand{\mvxi}{\boldsymbol{\xi}}
																																  
																																  \newcommand{\mvPsi}{\boldsymbol{\Psi}}

																																  \newcommand{\bD}{\mathbb{D}}

																																  \newcommand{\bN}{\mathbb{N}}
																																  \newcommand{\bR}{\mathbb{R}}
																																  \newcommand{\bS}{\mathbb{S}}
																																  
																																  \newcommand{\bZ}{\mathbb{Z}}

																																  \newcommand{\dD}{\mathds{D}}

																																  \DeclareMathOperator{\E}{\mathds{E}}
																																  \DeclareMathOperator{\pr}{\mathds{P}}
																																  
																																  \DeclareMathOperator{\var}{Var}
																																  \DeclareMathOperator{\cov}{Cov}

																																  \DeclareMathOperator{\argmin}{argmin}

																																  \usepackage{amsmath}

																																  \newtheorem{@assumption}{\sc Assumption}
																																  \newenvironment{assumption}{\begin{@assumption}\rm}{\end{@assumption}}

																																  \newcommand {\corr}{\qopname\relax n{\textrm{Corr}}}

																																  \newtheorem{remark}{Remark}[section]

																																  \newcommand{\s}{^{(s)}}

																																  \newcommand{\zhat}{\widehat{z}}
																																  \newcommand{\nhat}{\widehat{n}}
																																  \newcommand{\wb}{\bar}

\DeclareMathAlphabet\mathbfcal{OMS}{cmsy}{b}{n}
\usepackage{float}

\begin{document}

\sloppy
\title{Correlation networks, dynamic factor models \\ and  community detection
\footnote{The first and third authors were supported in part by the NSF grants DMS-2113662 and DMS-2134107. The second author was supported in part by NSF grant DMS-2134107.} 
}

\author{Shankar Bhamidi \quad \;  Dhruv Patel \quad \;  Vladas Pipiras \quad \; Guorong Wu \\ 
University of North Carolina}

\date{\today}
\maketitle

\begin{abstract}
A dynamic factor model with a mixture distribution of the loadings is introduced and studied for multivariate, possibly high-dimensional time series. The correlation matrix of the model exhibits a block structure, reminiscent of correlation patterns for many real multivariate time series. A standard $k$-means algorithm on the loadings estimated through principal components is used to cluster component time series into communities with accompanying bounds on the misclustering rate. This is one standard method of community detection applied to correlation matrices viewed as weighted networks. This work puts a mixture model, a dynamic factor model and network community detection in one interconnected framework. Performance of the proposed methodology is illustrated on simulated and real data. 
\end{abstract}

\section{Introduction}\label{c:intro}
Community detection methods for weighted networks have been applied extensively to correlation matrices of time series data \cite{gates:2016,macmahon:2015,masuda:2018,zhang:2005,worsley:2005}. Conceptually similar procedures have also been formulated based on various forms of hierarchical clustering of correlation matrices \cite{liu:2012}. This is often carried out in a way that is agnostic to an underlying data generating model (for which the correlation matrix is computed), with limited theoretical guarantees on the validity of community detection. In this work, we propose one natural model for the data generating process, define the notion of community explicitly, consider a standard community detection method and provide results for its performance under the model. 

The proposed model builds on a (dynamic) factor model for a $d$-vector stationary time series $\mvX_t = (X_{i,t})_{i=1,\ldots,d}$ driven by a low-dimensional $r$-vector stationary factor series $\mvf_t$, where $t$ refers to discrete time. The connection to communities is postulated through a $d\times r$ loading matrix $\mvLambda$ consisting of $d$ rows $\mvgl'_i$, with column vectors $\mvgl_i \in \bR^r$. (All vectors in this work are assumed to be column vectors and prime indicates transpose.) We assume that $\mvlambda_i$'s are drawn from a distribution $G_{\mvgl}$ on $\bR^r$ (in fact, a unit disc in $\bR^r$), and say that the model has $K$ communities when $G_{\mvgl}$ is a mixture of $K$ distributions $\{G_{k,\mvgl}\}_{k = 1,\ldots,K}$. For identifiability purposes, we require that the means of $G_{k,\mvgl}$ are distinct. More precise assumptions on $G_{\mvgl}$ are given in Section \ref{s:k-meansCDFM}. The component $i \in [d]:= \{1,\ldots,d\}$ is in community $k$ when $\mvgl_i$ is drawn from $G_{k,\mvgl}$. The exact model, called Community Dynamic Factor Model (CDFM), is defined in Section 2 below. 

We examine a number of issues for the CDFM: the block structure of the resulting correlation matrix, parametric mixing distributions $G_{k,\mvlambda}$, extensions to covariance matrices rather than correlation matrices, connections to other related constructions such as random dot product networks \cite{young:2007,athreya:2018} and reduced-rank Vector AutoRegressive (VAR) models. We also consider model estimation, including community detection methods, focusing on the standard spectral clustering through $k$-means, for which we apply a general theoretical result on the misclustering rate, obtained in \citet{patel:2022}, for the CDFM setting. 

The construction of the CDFM is motivated in part by the apparent block structure of correlation matrices for many multivariate time series. Several such matrices are examined in connection to the model, where we are particularly interested in the presence of mixing distributions (communities) and the dependence of clustering on separation of the mixing distributions. Applications presented in Section \ref{s:DS} below concern macroeconomic time series and fMRI data. 

The construction of the CDFM is conceptually simple and connected to other approaches. The loading matrix $\mvLambda$ will be estimated below through principal components of the correlation matrix. Thinking about mixtures and clustering of principle components is certainly not new, especially in the applied context \cite{jolliffe:2002}. On the other hand, correlation matrices can be viewed as weighted networks, amenable to network analyses. Our key contribution is that the CDFM allows us to connect a mixture model, a dynamic factor model (and principal components) and network community detection in one framework. 

It should be noted that the proposed framework and the results could also be cast in a multivariate, non-time series context, where time $t$ is replaced by index $n$, associated for example with different subjects, and where factors $\mvf_n$ are thought as i.i.d.\ across $n$. We work in the time series context for several reasons. Our own research interests aside, the stationary factors $\mvf_t$ do not need to be independent across time in the time series context. When $\mvf_t$ follow a Vector AutoRegressive (VAR) model, the CDFM is related to large VAR models with network structure, as we discuss in this work (Section \ref{s:COC}). Our applications are also for time series (Section \ref{s:DS}), and interesting future directions (e.g.\ change point detection) concern specifically data with temporal ordering.

The rest of the paper is structured as follows. The CDFM is formulated and studied in Section \ref{s:CDFM}. Estimation and community detection for the CDFM are considered in Section \ref{s:E&CD}.  Simulations and applications to real data can be found in Section \ref{s:DS}. Connections to other constructions are discussed in Section \ref{s:COC}. We conclude in Section 6 and technical assumptions are stated in Appendix \ref{Appendix-assumps}. 


\section{Community Dynamic Factor Model}\label{s:CDFM}
\subsection{Model Formulation}\label{s:MF}

Let $[d] = \set{1,2,\ldots, d}$ denote the node set.  Let $\set{\mvX_t: t\in \bZ}$ be a $d$-dimensional stationary time series following a dynamic factor model (DFM), and more specifically its so-called static form \cite{stock:2016} given by
\begin{equation}\label{DFM}
\mvX_t = \mvLambda \mvf_t + \mvgee_t, \;\;\; t \in \bZ,
\end{equation}
where $\{\mvf_t: t\in \bZ\}$ is an $r$-dimensional stationary time series, $\mvLambda = (\mvlambda_i)_{i=1,\ldots,d}$ is a $d\times r$ loadings matrix with $d$ row vectors $\mvgl_i'$ and $\mvgl_i \in \bR^r$, and $\mvgee_t$ are the error terms with $\E \mvgee_t = \mvzero$ and $\E \mvgee_t \mvgee_t' = \mvgS_{\epsilon}$. The factors $\{\mvf_t\}$ can be dependent across time and are assumed to have zero mean. The errors $\{\mvgee_t\}$ and factors $\{\mvf_t\}$ are assumed independent with further assumptions stated below in connection to the theoretical results. As common in the DFM literature, we mitigate the issue of non-identifiability of the DFM by assuming 
\begin{equation}\label{a-factor_cov}
\E \mvf_t \mvf'_t = \mvgS_f = \mvI_r,
\end{equation}
so that $\mvgL,\mvf_t$ are now identified up to an orthogonal transformation (specified in Section \ref{s:LFest}). We shall also discuss the case when \eqref{a-factor_cov} is not assumed in Section \ref{oblique factors}.

For the loadings, fix $K\geq 1$, distinct probability measures $\set{G_{k,\mvlambda}:k \in [K]}$ on $\bR^r$ and a probability mass function $\vp = \set{p_k:k \in [K]}$. Consider the mixture distribution

\begin{equation}\label{mixture_model}
	G_{\mvlambda} = \sum_{k=1}^K p_k G_{k,\mvlambda}. 
\end{equation}
The loading rows $\mvgl'_1,\ldots,\mvgl'_d$ are assumed to be independent and  
\begin{equation}\label{Loading_matrix}
\mvlambda_i \sim G_{\mvlambda},
\end{equation}
that is, $\mvlambda_i$'s are drawn independently from $G_{\mvlambda}$. 
\begin{defn}\label{def:cdfm}
The community dynamic factor model (CDFM) is defined as the DFM \eqref{DFM} with loadings given by \eqref{Loading_matrix}. 
\end{defn}
Further technical assumptions are specified in Section \ref{s:k-meansCDFM};  for the moment, the reader can think of the mixing distributions $\{G_{k,\mvgl}\}_{k\in[K]}$ as point masses, Gaussian distributions or other unimodal distributions with distinct location parameters. We also assume that $\mvgl_i$'s are drawn and fixed. In particular, the expectation sign $\E$ throughout stands for the conditional expectation given $\{\mvgl_i\}_{i\in[d]}$, and $\E_{G_{\mvgl}}$ for the expectation with respect to the randomness in the loadings $\{\mvgl_i\}_{i\in[d]}$ .  

Assume that the component series are normalized, so that
\begin{equation}\label{secondmomentassump}
\E(X_{i,t}^2) = 1, \qquad \forall i\in [d], \quad t\in \bZ. 
\end{equation}
Since 
\begin{equation}\label{lambda'lambda restriction}
\E(X^2_{i,t}) = \mvlambda_i'\mvlambda_i + (\mvgS_{\gee})_{ii} = 1,
\end{equation}
we must have that $\| \mvlambda_i\|_2 \leq 1$. This constraints the mixing distributions $G_{k,\mvlambda}$  to the unit disc $\dD^{r-1} : = \{ \mvgl \in \bR^r : \|\mvgl \|_2 \leq 1 \}$. 
Furthermore, this assumption enforces that the correlation matrix and the covariance matrix (conditional on the loadings) are the same and given by 
\begin{equation}\label{covariance_X}
\mvgS_X = \E\mvX_t\mvX_t' = \mvLambda \E(\mvf_t\mvf_t') \mvLambda' + \mvgS_{\gee} = \mvLambda \mvLambda' + \mvgS_{\gee}
\end{equation}
by using \eqref{a-factor_cov}. Write
\begin{equation}\label{sample cov_X}
\wh{\mvgS}_X = \dfrac{1}{T}\sum_{t=1}^T \mvX_t \mvX_t'
\end{equation}
for the sample covariance matrix, where the sample mean is not subtracted for simplicity. We also ignore the issue of standardization of data. The sample covariance will not be used until Section \ref{s:E&CD}. 

The individual components of the time series are driven by the loadings drawn from $K$ mixing distributions. Naturally, one can group the $d$ components into $K$ communities given by the respective distributions of the rows of $\mvLambda$.
\begin{defn}\label{def:communities}
Let $\mvZ = (Z_{i,k} : i\in [d]; k \in [K])$ be the $d\times K$ membership matrix given by $Z_{i,k} = 1$ if $\mvlambda_i$ is drawn from $G_{k,\mvlambda}$ and $ = 0$ otherwise. Let $z: [d] \to [K]$ be the membership function given by $z[i] = j$ if $Z_{i,j} = 1$ and $= 0$ otherwise.  Communities for the components $i \in [d]$ are defined as $\cC_k = \{ i : Z_{i,k} = 1 \}$, $k \in [K]$. Define $n_k = |\cC_k|$ to be the size of community $k \in [K]$.
\end{defn}


\subsection{Block Structure of Correlation Matrices}\label{s:block}

A feature of the CDFM is that its correlation matrix will have a block structure on ``average." In practice, possibly after reordering by communities, sample correlation matrices do often exhibit block structure. This suggests our model as a candidate to capture this phenomenon. 

Indeed, assume that the $d$ component series of $\mvX_t$ are ordered by community membership so that 
\begin{equation}\label{eq:Z}
\mvZ = \begin{bmatrix}
\mvone_{n_1} & \mvzero & \ldots  & \mvzero \\
 \mvzero & \mvone_{n_2} & \cdots &  \mvzero \\
 \vdots & \vdots & \ddots & \vdots\\ 
 \mvzero & \mvzero & \cdots & \mvone_{n_K}
\end{bmatrix},
\end{equation} 
where $\mvone_n \in \bR^n$ is a vector of  $1$'s and $\mvzero$'s are vectors of  $0$'s of the appropriate dimensions. The community structure in $\mvX_t$ translates into the block structure of the component $\mvLambda \mvLambda'$ of the correlation matrix on ``average" as follows. Under the mixture measure, given $\mvZ$ in \eqref{eq:Z}, the expected $\mvLambda \mvLambda'$ is a block matrix given by
\begin{equation}\label{eq:E_LambdaLambda}
\E_{G_{\mvlambda}}(\mvLambda \mvLambda'| \mvZ) = \mvZ \begin{bmatrix} \mvmu_1' \\ \vdots \\ \mvmu_K' \end{bmatrix}
\begin{bmatrix} \mvmu_1 & \ldots & \mvmu_K \end{bmatrix} \mvZ'  = [ (\mvmu_k' \mvmu_l)  \mvJ_{n_k,n_l}]_{k,l \in [K]},
\end{equation} 
where $\mvJ_{n_k,n_l}$ is a $n_k \times n_l$ matrix of $1$'s and
\begin{equation}\label{def:mixture_means}
\mvmu_k = \E_{G_{k,\mvgl}}(\mvgl), \quad k \in [K],
\end{equation}
are the means of the mixing distributions. The main diagonal blocks $\{(\mvgm_k'\mvgm_k)\mvJ_{n_k,n_k} \}$ of \eqref{eq:E_LambdaLambda} are characterized by the inner products of the means $\{\mvmu_k' \mvmu_k\}$. If the inner products are unique, then the main diagonal structure is sufficient for community identifiability. Although this is sufficient, it is not necessary. In fact, even if for some $k \neq k'$, we have that $\mvmu_k' \mvmu_k= \mvmu_{k'}'\mvmu_{k'}$, as long as there exists an $l$ such that $\mvmu_k'\mvmu_l \neq \mvmu_{k'}' \mvmu_l$, we will still see the community block structure in $\E_{G_{\mvlambda}}(\mvLambda \mvLambda'| \mvZ)$.

Two other observations are worth making. First, the only restriction is $\|\mvmu_k\|_2 \leq 1$. Second, what determines the number of blocks is $K$ (the number of mixing distributions), not $r$ (the number of factors). That is, one can have a block matrix even for one factor $r=1$ as in examples in Section \ref{s:Ex-CDFM}.


\subsection{Case of Covariance Matrices}\label{s:cov}

We focused above on correlation matrices assuming that $\E(X_{i,t}^2) = 1$. Without this assumption the covariance matrices $\mvgS_{X}$ can, in principle, be treated similarly. In fact, the discussion simplifies in that the mixing distributions can now be on $\bR^r$ (instead of $\bD^{r-1})$.  A popular Gaussian mixture distribution can be considered for $G_{\mvlambda}$. 

However, this model for $\mvgS_{X}$ presents interpretation issues: is the block (community) structure now driven by the variances or the correlations or some combination of the two? For example, when $r=1$, $K=2$ and $\mvgS_{\gee} = a\mvI_d$ for some $a > 0$, $\var(X_{i,t}) = \gl_i^2 + a$, where we take $\gl_i>0$. If $a$ is small relative to $\{\gl_i\}_{i\in [d]}$, especially for large $\gl_i$'s, $\var(X_{i,t}) \approx \gl_i^2$. So the variances contain the community information. However, the correlations,
\begin{equation}
\corr(X_{i,t},X_{j,t}) = \dfrac{\cov(X_{i,t},X_{j,t})}{\var(X_{i,t})^{1/2}\var(X_{j,t})^{1/2}} \approx \dfrac{\gl_i\gl_j + a}{\gl_i\gl_j} \approx 1,
\end{equation}
contain little community information. If $a$ is comparable to $\{\gl_i\}_{i\in [d]}$, then the correlations also contain the community structure. 

For this interpretation reason and also since sample correlation matrices are usually of interest in practice, we advocate to work with correlation matrices. The community structure on variances could also be of interest, and can be studied through other means.

Finally, we also note that the relationship between covariance and correlation matrices is akin to the relationship between the adjacency matrices of networks and their degree corrected counterparts. Let $\mvA = (A_{i,j})_{i,j\in[d]}$, with $A_{i,j} >0$ for simplicity, be the adjacency matrix of a weighted network. The degree corrected adjacency matrix is 
\begin{equation*}
\mvD^{-1/2} \mvA \mvD^{-1/2} = \bigg ( \dfrac{A_{i,j}}{D_{i,i}^{1/2}D_{j,j}^{1/2}} \bigg)_{i,j \in [d]},
\end{equation*}
where $D_{i,i} = \sum_{j \neq i} A_{i,j}$ is the degree of node $i$ and $\mvD = diag (D_{1,1},\ldots ,D_{d,d})$ \citep{karrer:2011}. For correlation matrices, the variance plays the role of the degree. Furthermore, as the properties of the covariance matrix can be dictated by the variances according to the discussion above, the same holds for node degrees in adjacency matrices. For example, without degree correction, it is known that the distribution of the eigenvalues of the adjacency matrix follow that of the degrees, even if there is an underlying community structure \citep{zhan:2010}.


\subsection{Case of Non-Orthogonal Factors and Effect of Rotations}\label{oblique factors}

We assumed in \eqref{a-factor_cov} that the factors are orthogonal, motivated in part by the principal component analysis (PCA) used below, which yields this orthogonality. If \eqref{a-factor_cov} is not satisfied, the covariance of $\mvX$ is given by $\mvgS_{X} = \mvgL \mvgS_{f}\mvgL' + \mvgS_{\gee}$.
Letting $\mvgS_{f} = \mvQ_f \mvD_f \mvQ_f'$ be the eigendecomposition with orthogonal $\mvQ_f$ and diagonal $\mvD_f$, $\wb\mvgL = \mvgL  \mvQ_f \mvD_f^{1/2}$ and $\wb\mvf_t = \mvD_f^{-1/2} \mvQ_f' \mvf_t$, the DFM \eqref{DFM} can be written as $\mvX_t = \wb\mvgL \wb\mvf_t + \mvgee_t$ with $\E \wb\mvf_t \wb\mvf_t' = \mvI_r$ and $\mvgS_{X} = \wb\mvgL \wb\mvgL' + \mvgS_{\gee}$. Note that the transformation $\mvQ_f \mvD_f^{1/2}$ does not necessarily preserve angles. See Example \ref{Example1} for such a case. 

Another aspect of the above transformation to keep in mind is that $\mvgL$ can be sparser than $\wb\mvgL$ and thus potentially more interpretable (or vice versa). It should be noted that the issue of sparsity of $\mvgL$ is pertinent for orthogonal factors as well. The orthogonality of factors is preserved under orthogonal transformation (e.g.\ rotation), but the resulting $\mvgL$ can indeed become sparser. This is the basis of the traditional VARIMAX procedure (see \citet{rohe:2020} for a more modern take). More generally, estimation of sparse loadings is the objective of sparse PCA and related approaches \cite{guerra-urzola:2021}.  The focus of this work will be on orthogonal factors and for identifiability purposes, we shall focus on a specific rotation such that 
\begin{equation}\label{diagonal_lambda}
\mvgL'\mvgL = \mvP,
\end{equation}
for some diagonal matrix $\mvP$. Indeed, the DFM is identifiable when imposing the constraints \eqref{a-factor_cov}, \eqref{diagonal_lambda}, and that the entries of $\mvP$ are distinct. 

\subsection{Parametric Families of Mixing Distributions}\label{s:Families}

The mixture model with the assumptions above restricts the distributions $\{G_{k,\mvgl}\}$ to the unit disc $\bD^{r-1}$. In this section, we discuss some parametric families of mixing distributions on $\bD^{r-1}$. The discussion below concerns properties of a typical component of the mixture, so we drop $k$ from the notation for simplicity. We also replace $\mvlambda\in \dD^{r-1}$ by a general random variable $\mvY_D\in \dD^{r-1}$ in the discussion below.

\subsubsection{Point Mass}
The simplest family of mixing distributions is the point mass family where each mixing distribution has all of its mass on $\mvgm_k$, i.e.\ $\mvY_D \sim \gd_{\{\mvgm_k\}}$. As long as the points $\{\mvgm_k\}$ are distinct, the mixture is identifiable. Point mass mixtures are a common choice of mixing distributions as they allow constant values within communities. For example, in the Random Dot Product Graph (RDPG) literature, it is often assumed that the latent attributes of nodes from a single community are constant. More details on RDPGs and their connections to the CDFM are provided in Section \ref{s:RDPG}.

\subsubsection{Projected Normal Distribution}\label{s:PND}

Let $\mvY \sim \cN_r(\mvgm_{Y}, \mvI_r)$, the $r$-dimensional normal distribution with mean $\mvgm_Y$ and covariance $\mvI_r$, and define $\widetilde{\mvY} = {\mvY}/{\|\mvY\|_2}$. Then, $\wt{\mvY}$ is distributed as a projected normal distribution (PND), $\cP\cN_r(\mvgm_Y,\mvI_r)$. The PND $ \widetilde{\mvY}$ is a unimodal and
symmetric distribution on $\bS^{r-1} := \{\mvgl \in \bR^r: \|\mvgl\|_2 = 1\}$ with mean ${\mvgm_Y}/{\|\mvgm_Y\|_2}$. The PND has identifiability issues: for any constant $c >0$, $\widetilde{\mvY}$ has the same distribution as $(\widetilde{c\mvY})$. 

To construct distributions on $\bD^{r-1}$, let $\mvY_D = \widetilde{\mvY}\cB$ where $\cB \sim Beta(a,b)$ with $a, b > 0$ is independent of $\widetilde{\mvY}$. One may take $\cB$ to follow other parametric unimodal distributions on $[0,1]$. Then, $\mvY_D$ follows a unimodal distribution on $\bD^{r-1}$. See Example \ref{ex:PND-Beta} for a numerical illustration of a model from this class. 
 
\subsubsection{Restricted Normal Distribution}\label{s:RND}

Let $\mvY \sim \cN_r (\mvgm_Y, \mvgS_Y)$ with positive definite $\mvgS_Y $. The restriction of $\mvY$ to the unit disc, denoted $\mvY_D$, has a probability distribution function (pdf) given by 
\begin{equation}\label{rnd-density}
f_{\mvY_D}(\mvy) = \dfrac{f_{\mvY}(\mvy)\cI ( \mvy \in \bD^{r-1})}{\pr (\mvY \in \bD^{r-1})},
\end{equation}
where $f_{\mvY}$ is the pdf of $\mvY$. To calculate $\pr (\mvY \in \bD^{r-1})$, one notes that $\|\mvY\|^2_2$ is of quadratic form and hence the distribution can be calculated explicitly as follows. Let $\mvgS_Y = \mvQ_Y \mvD_Y \mvQ'_Y$ be the eigendecomposition with orthogonal $\mvQ_Y$ and diagonal $\mvD_Y$. Then, $\|\mvY\|^2_2 = \mvY' \mvY$ has a generalized chi-squared distribution,
\begin{equation}
\mvY'\mvY \sim \sum_{j=1}^r (\mvD_Y)_{j,j} \; \chi^2(b^2_j),
\end{equation}
where $ \chi^2(b^2_j)$ are independent non-central chi-squared distributions with the non-centrality parameters given by the vector $\mvb = \mvQ_Y'\mvgS_Y^{-{1}/{2}}\mvgm_Y = (b_j)_{j\in[r]}$ \cite{mathai:1992}. Following \citet{imhof:1961}, when the eigenvalues of $\mvgS_Y$ are assumed to be unique, the CDF of $\mvY'\mvY$ can eb expressed as
\begin{align}
F_{\mvY'\mvY} (y) &= \dfrac{1}{2} - \dfrac{1}{\pi} \int_0^{\infty} \dfrac{\sin(\gt(y,u))}{u\gr(u)} du,  \label{imhof-method}
\end{align}
where
\begin{align}
\gt(y,u) & = \dfrac{1}{2} \sum_{j=1}^r (\tan^{-1}(\gd_j u) + b_j^2\gd_ju(1+\gd_j^2u^2)^{-1}) - \dfrac{uy}{2}, \\
\gr(u) &= \prod_{j=1}^{r} (1+b_j^2u^2)^{\frac{1}{4}} \exp\bigg ( \dfrac{1}{2}\sum_{l=1}^r \dfrac{(b_l\gd_lu)^2}{1+\gd_l^2u^2} \bigg )
\end{align}
and $\mvgd = (\gd_1 ,\ldots, \gd_r) = diag(\mvD_Y)$. The expression \eqref{imhof-method} can be used to compute $\pr (\mvY \in \bD^{r-1}) = F_{\mvY'\mvY} (1)$ in \eqref{rnd-density}.


\subsection{Examples of CDFMs}\label{s:Ex-CDFM}
We illustrate the various concepts introduced above through several examples of CDFMs.

\begin{figure}[t]
\centering
\includegraphics[width=\textwidth,height = 2in]{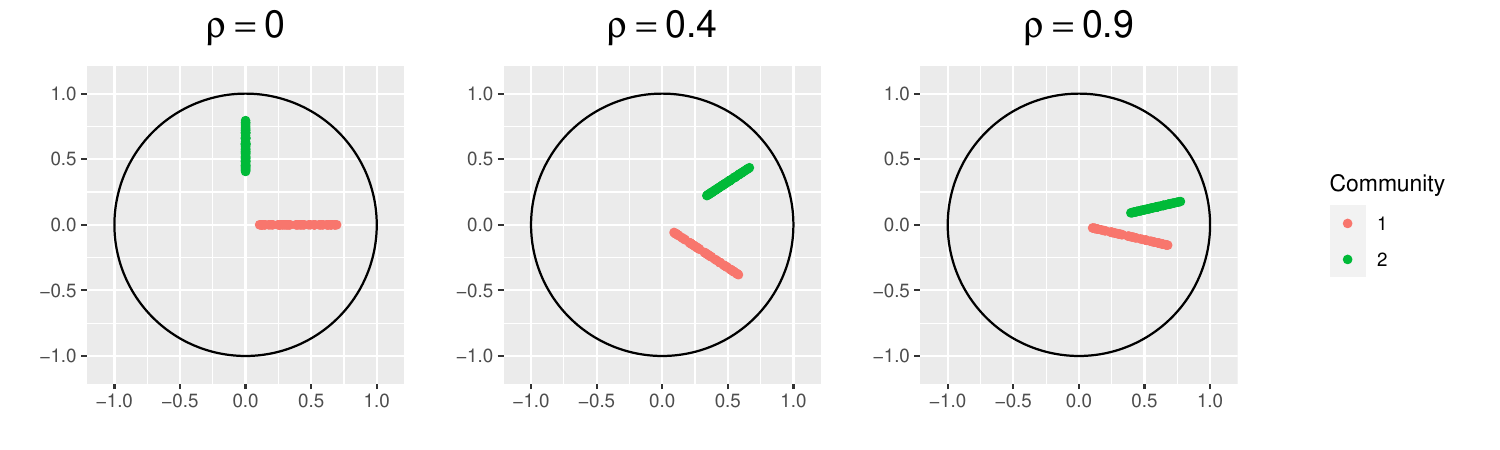}

\caption{Plotted are the rows of $\mvgL$ in \eqref{Ex1-rewritten lambda} with $n_1 = n_2 = 60$. As $\rho$ increases, the separability between the two communities decreases. }\label{gates-example}
\end{figure}

\begin{ex}\label{Example1}
Consider $r = K = 2$. Fix $\gr$ with $|\gr| <1$ and   
\begin{equation}\label{Ex1-model}
\mvLambda = \begin{bmatrix}
\mvgL_1 & \mvzero \\ \mvzero & \mvgL_2
\end{bmatrix} = 
\begin{bmatrix}
\lambda_{1,1} & 0 \\ \vdots & \vdots \\ \lambda_{n_1,1} & 0 \\ 0 & \lambda_{n_1+1,2} \\ \vdots & \vdots \\ 0 & \lambda_{n_1+n_2,2}
\end{bmatrix} , \quad  \mvgS_f = \begin{bmatrix}
1 & \rho \\ \rho & 1  
\end{bmatrix},
\end{equation}
where $\lambda_{i,j} \sim \cU(a,b)$, the uniform distribution on $(a,b)$, independent across indices with $0<a<b<1$. Thus, we assume that $\mvlambda_i$, $i=1,\ldots,n_1$, for community 1 are drawn from $G_{1,\mvlambda} = \cU(a,b)\times \delta_{\{0\}}$ and $\mvlambda_i$, $i= n_1+1,\ldots,n_2,$ for community 2 from $G_{2,\mvlambda} = \delta_{\{0\}} \times \cU(a,b)$. The cross correlation term $\rho$ controls the dependence between the strengths of nodes in communities 1 and 2. A distributionally equivalent representation is given by moving $\rho$ from $\mvgS_f$ to $\mvLambda$ as
\begin{equation}\label{Ex1-rewritten lambda}
\mvLambda = \frac{1}{2}\begin{bmatrix}
\mvgL_1(\sqrt{1+\rho} + \sqrt{1 - \rho}) & \mvgL_1(\sqrt{1+\rho} - \sqrt{1 - \rho}) \\ \mvgL_2(\sqrt{1+\rho} - \sqrt{1 - \rho}) & \mvgL_2(\sqrt{1+\rho} + \sqrt{1 - \rho})
\end{bmatrix}, \quad  \mvgS_f = \begin{bmatrix}
1 & 0 \\ 0 & 1 \end{bmatrix}.
\end{equation}
Now, the dependence of the factors is put on the mixing distributions as $G_{1,\mvlambda} = \cU(a,b)\delta_{(\sqrt{1+\rho} + \sqrt{1 - \rho},\sqrt{1+\rho} - \sqrt{1 - \rho})}$ and $G_{2,\mvlambda} = \cU(a,b)\delta_{(\sqrt{1+\rho} - \sqrt{1 - \rho},\sqrt{1+\rho} + \sqrt{1 - \rho})}$. The effects of $\gr$ can be seen in Figure \ref{gates-example}. Note that as $\rho$ increases, we lose separability of the two communities. The model \eqref{Ex1-model} appears in \citet{gates:2016}.
\end{ex}

The number of factors $r$ does not need to equal the number of communities $K$. The following are simple examples with $r \neq K$. 

\begin{ex}\label{Uniform Circle Example}
Let $K = 1$ and $r=2$. Take $\mvlambda_i = m_i (\cos(\theta_i),\sin(\theta_i))$ with i.i.d.\ $\theta_i \sim\cU(0,2\pi]$ and $m_i \sim \cU(0,1)$. Equivalently, $G_{1,\mvlambda} = U \frac{Q}{\|Q\|_2}$ with $Q \sim \cN_2(\mvzero,\mvI_2)$ and $U \sim \cU(0,1)$. This is a similar construction to the parametric family defined in Section \ref{s:PND}. 

Consider another setting where $K =2$ and $r=1$, let $G_{1,\mvlambda} = \delta_{\{1\}}$ and $G_{2,\mvlambda} = \delta_{\{-1\}}$. Then each $\mvlambda_i$ is either $1$ or $-1$ depending on its community assignment. 
\end{ex}

Let $\mvY \sim \cN_r(\mvgm_{Y}, \mvI_r)$, the $r$-dimensional normal distribution with mean $\mvgm_Y$ and covariance $\mvI_r$, and define $\widetilde{\mvY} = {\mvY}/{\|\mvY\|_2}$. Then, $\wt{\mvY}$ is distributed as a projected normal distribution (PND), $\cP\cN_r(\mvgm_Y,\mvI_r)$. The PND $ \widetilde{\mvY}$ is a unimodal and
symmetric distribution on $\bS^{r-1} := \{\mvgl \in \bR^r: \|\mvgl\|_2 = 1\}$ with mean ${\mvgm_Y}/{\|\mvgm_Y\|_2}$. The PND has identifiability issues: for any constant $c >0$, $\widetilde{\mvY}$ has the same distribution as $(\widetilde{c\mvY})$. 

To construct distributions on $\bD^{r-1}$, let $\mvY_D = \widetilde{\mvY}\cB$ where $\cB \sim Beta(a,b)$ with $a, b > 0$ is independent of $\widetilde{\mvY}$. One may take $\cB$ to follow other parametric unimodal distributions on $[0,1]$. Then, $\mvY_D$ follows a unimodal distribution on $\bD^{r-1}$. See Example \ref{ex:PND-Beta} for a numerical illustration of a model from this class.

\begin{figure}[t]
\centering
\includegraphics[width=\textwidth,height = 2in]{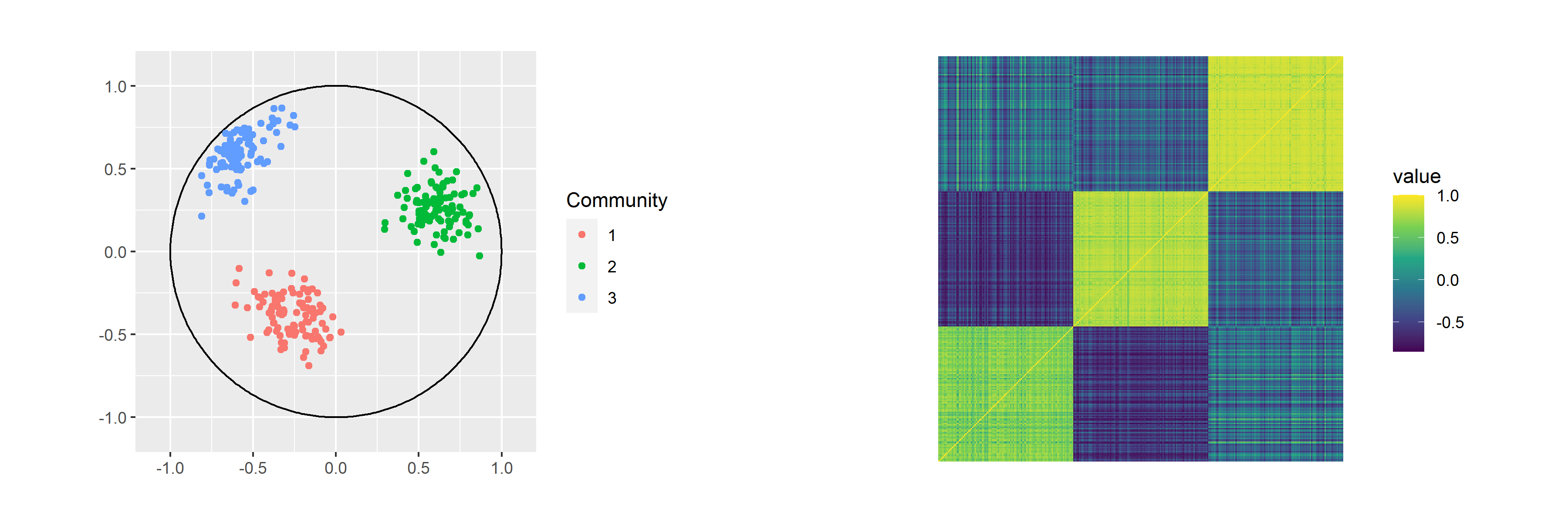}
\caption{Left: Vectors sampled from $G_{1,\mvlambda}$ (Red), $G_{2,\mvlambda}$ (Green), $G_{3,\mvlambda}$ (Blue). Right: Heatmap of the associated correlation matrix $\mvgS_{X}$.}\label{PND-Beta-plots}
\end{figure}

\begin{ex}\label{ex:PND-Beta}
Consider the PND-Beta distribution as described in Section \ref{s:PND}. Let $\mvY_k \sim \cN_r(\mvgm_{k,\mvY}, \mvI_r)$ and $\cB_k \sim Beta(a_k,b_k)$, $a_k, b_k > 0$, for $k \in [K]$ with $\{\mvY_k\}$ independent of $\{\cB_k\}$. Define $\wt\mvY_k = {\mvY_k}/{\|\mvY_k\|_2}$ and 
\begin{equation}\label{Beta*PN Mixture}
G_{k,\mvlambda} \sim  \wt\mvY_k \cB_k.
\end{equation}
So, $G_{k,\mvlambda}$ is a distribution on $\bD^{r-1}$ such that for any vector sampled from $G_{k,\mvlambda}$, the angle is determined by the normal distribution $\mvY_k$ and the magnitude is determined by $\cB_k$. As an example, take $r = 2$ and $K=3$. Set $(a_1,b_1) = (10,10),(a_2,b_2) = (10,5),(a_3,b_3) = (20,4)$, $ \mvgm_{1,\mvY} = (2,3)',\mvgm_{2,\mvY} =(-5,-2)'$, and $\mvgm_{3,\mvY} =(4,-4)'$. We sample 100 vectors from each of $G_{k,\mvgl}$ for $k =1,2,3$. The plots of $\{\mvgl_i\}$ and $\mvgS_{X}$ are shown in Figure \ref{PND-Beta-plots}. Note that, even though the main block diagonal does not clearly show three communities (but rather two communities), the off diagonal blocks show three communities. 
\end{ex}

\begin{remark} In view of \eqref{lambda'lambda restriction}, the closer $\mvgl_i$ is to the unit circle, the stronger the ``signal" $\mvf_t$ is in the component series $X_{i,t}$ (equivalently, the weaker the noise $\gee_{i,t}$). This also means that the closer $\mvgl_i$'s from a community are to the unit circle, the stronger the correlations will be among the component series $X_{i,t}$'s. This can be seen in Figure \ref{PND-Beta-plots}, where community 3 (blue) is closest to the unit circle and has the correlation block (top right) with the largest values. This perspective should also be kept in mind with estimated loadings in practice, as in the data illustrations in Section \ref{s:DS}.
\end{remark}


\section{Estimation and Community Detection}\label{s:E&CD}

We discuss here estimation questions for the CDFM, including estimation of communities. We work in the regime where both the dimension $d$ and the sample size $T$ can be large. We assume throughout that the number of factors $r$ and the number of communities $K$ are known. In Section \ref{s:chooseK} below, we discuss one method to choose $K$ in practice and refer the reader to \citet{stock:2016} for methods to pick $r$. 

\subsection{Loading and Factor Estimation}\label{s:LFest}
We recall here several known results on estimation of loading matrix $\mvgL$ and factors $\mvf_t$, as needed for subsequent community detection. The loadings and factors are estimated through a standard PCA approach as follows. Let $\wh\mvgS_{X} = \wh\mvQ \wh\mvD \wh\mvQ'$
be the eigendecomposition of the sample covariance matrix \eqref{sample cov_X} where $\wh\mvQ$ consists of the orthogonal eigenvectors and the diagonal matrix $\wh\mvD$ consists of the respective eigenvalues, in decreasing order. Let $\wh\mvQ_r$ be the $d\times r$ matrix of the first $r$ eigenvectors of $\wh\mvgS_X$ associated with the $r$ largest eigenvalues forming a diagonal matrix $\wh\mvD_r$. The PCA estimators are defined as 
\begin{equation}\label{PCA estimates}
\wh\mvgL = \wh\mvQ_r \wh\mvD_r^{\frac{1}{2}}, \quad \wh\mvf_t = \wh\mvD_r^{-1}\wh\mvgL'\mvX_t.
\end{equation}
Note that, by construction,
\begin{equation}\label{PCA constraints}
\wh\mvgL'\wh\mvgL = \wh\mvD_r , \quad \dfrac{1}{T}\sum_{t=1}^T \wh\mvf_t\wh\mvf_t' = \mvI_r.
\end{equation}

The constraints \eqref{PCA constraints} allow one to identify the limits of $\wh\mvgL$ and $\wh\mvf_t$ in terms of $\mvgL$ and $\mvf_t$. More specifically, let 
\begin{equation}\label{lambda'lambda}
\mvgL'\mvgL = \mvQ_{\gl} \mvD_{\gl} \mvQ_{\gl}'
\end{equation}
be the eigendecomposition of $\mvgL'\mvgL$. (Do not confuse $\mvQ_{\gl}$ and $\wh\mvQ_r$; even their dimensions are different.) Then, $\mvgL\mvf_t = (\mvgL\mvQ_{\gl})(\mvQ_{\gl}'\mvf_t)$. Since $(\mvgL\mvQ_{\gl})'\mvgL\mvQ_{\gl} = \mvD_{\gl}$ is diagonal as is the first relation in \eqref{PCA constraints}, under suitable assumptions (\citet{bai:2008} and \citet{doz:2012}), as $d,T \to \infty$,
\begin{equation}\label{convergence of ests}
\|\wh\mvgL - \mvgL^{(0)}\|_* \to 0, \quad \wh\mvf_t \stackrel{p}{\to} \mvf_t^{(0)},
\end{equation}
where $\|\cdot\|_*$ is the spectral (operator) norm, $\mvgL^{(0)} := \mvgL\mvQ_{\gl}$, and $\mvf_t := \mvQ_{\gl}' \mvf_t $. Strictly speaking, the eigenvalues in $\mvD_{\gl}$ need to be different for this identifiability (see also \eqref{diagonal_lambda}) and one of the key assumptions is that of {\it strong factors}, essentially saying that the eigenvalues of $\mvgL'\mvgL$ need to be of the order $d$, which we express as
\begin{equation}\label{factors-strong}
	\mvgL'\mvgL \asymp d.
\end{equation}
\citet{bai:2008} and \citet{doz:2012} provide convergence rates for \eqref{convergence of ests} as well. 

For our results on community detection, we shall need the following stronger convergence result. 
\begin{prop}{[\citet{uematsu:2019}, Lemma 6]}\label{weak-factor-thm}
Let $v >0$ be an arbitrary constant. Under suitable assumptions on the DFM \eqref{DFM}, with probability at least $1 - O((d \vee T)^{-v})$,
\begin{equation}\label{lambda max bound}
\|\wh\mvgL - \mvgL^{(0)} \|_{\max} \leq C \bigg ( \dfrac{\log(d \vee T)}{T} \bigg)^{\frac{1}{2}}
\end{equation}
for some $C>0$, where $d \vee T = \max\{d,T\}$. 
\end{prop}
The detailed assumptions required for Proposition \ref{weak-factor-thm} are given and discussed in Appendix \ref{Appendix-assumps}. In fact, the above result is proven not only for strong factors, but also for {\it weak factors} such that, by using the notation in \eqref{factors-strong},
\begin{equation}\label{factors-weak}
	\mvgL'\mvgL \asymp d^\alpha
\end{equation}
for $\alpha\in (0,1)$. (Additional assumptions on $\alpha$, $T$ and $d$ need to be made as well.)


\subsection{Community Detection Through $k$-Means}
By \eqref{convergence of ests} and \eqref{lambda max bound}, PCA estimates the rotated loadings matrix $\mvgL^{(0)}$. Rotation preserves the mixture structure \eqref{mixture_model}, with each mixing distribution $G_{k,\mvgl}$ undergoing the same rotation. To simplify the notation, we shall drop the superscript from $\mvgL^{(0)}$ and simply write $\mvgL$ for the loading matrix estimated through PCA. 

We are interested here in estimating the communities $\cC_k$ in Definition \ref{def:communities}. With known $\mvgL$ (or $\mvgl_i$'s), a commonly used algorithm for clustering is $k$-means, which we recall in Section \ref{s:K-means} below. Theoretical results on misclassification rate are derived in \citet{patel:2022} by considering estimated loadings as perturbations of the original loadings under the assumption that the mixing distributions are well-separated. In Section \ref{s:k-meansCDFM}, we apply one of these results to the case where $\wh\mvgl_i$'s are thought of as perturbations of $\mvgl_i$'s of magnitude at most the bound in \eqref{lambda max bound}. Throughout we assume $K$ is known, deferring the reader to Section \ref{s:chooseK} for methodology to pick $K$.

The terms ``cluster'' and ``community'' will be used interchangeably, especially in connection to $k$-means, and should be understood as such.

\subsubsection{$k$-Means Algorithm}\label{s:K-means}

The $k$-means clustering technique aims to cluster a set of $d$ points $\{\mvx_i \in \bR^r$: $i \in [d] \}$ into $K$ clusters (communities) $\wh\cC_1, \ldots ,\wh\cC_K$, by minimizing 
\begin{equation}\label{4-def-kmeans}
\min_{\wh\mvC,\wh z} \dfrac{1}{d} \sum_{i=1}^d \|\mvx_i - \wh\mvm_{\wh z(i)} \|_2^2,
\end{equation}
where $\wh\mvC = \{\wh\cC_1,\ldots,\wh\cC_K\}$ is referred to as an estimated clustering, $\wh z:[d] \to [K]$ denotes an estimated cluster assignment (i.e.\ $\wh z(i) = k $ if $i \in \wh\cC_k$) and $\wh\mvm_k$ is the mean of the points in $\wh \cC_k$.

The $k$-means optimization problem is NP-hard \cite{mahajan:2009} and is tackled by a class of algorithms which attempt to approximately minimize \eqref{4-def-kmeans} by initializing a set of candidate means, then iterating between updating cluster assignments and updating cluster means. One of the most popular and simplest algorithms is Lloyd's algorithm \cite{lloyd:1982}, often referred to as the standard or naive $k$-means. The updating step in Lloyd's algorithm is as follows. Let $\{\wh\mvm^{(s)}_k\}$ be the means of the cluster assignments at the $s$-th iteration. Then, the cluster assignment of $\mvx_i$ at the $(s+1)$-th iteration is given by 
\begin{equation}
\wh z^{(s+1)}(i) = \argmin_{k \in [K]} \|\mvx_i - \wh\mvm^{(s)}_k \|_2.
\end{equation} 
Lloyd's algorithm converges to a local minimum but is reliant on good initialization to achieve a global minimum. Lloyd's algorithm is often initialized as follows. Sample $\wh\mvm_1^{(1)}$ randomly from $\{\mvx_i\}_{i\in[d]}$. For the remaining points, calculate the distance squared to the closest initialized mean and sample $\wh\mvm_2^{(1)}$ from the remaining points with probability proportional to this squared distance. Repeat the procedure until all $K$ means have been initialized. Lloyd's algorithm initialized in such a way is referred to as the $k$-means++ algorithm \citep{arthur:2007}. For an overview of the $k$-means algorithm and its variants, see \citet{wierzchon:2018}.


\subsubsection{$k$-Means for CDFM}\label{s:k-meansCDFM}

To estimate the communities in the CDFM, we apply $k$-means to the obtained PCA estimates $\{\wh\mvgl_i\}_{i\in[d]}$. This setting differs from most $k$-means settings in the mixture model literature \cite{mcnicholas:2016} as we are estimating, not observing, the draws from the mixture. We will focus on the performance of Lloyd's algorithm on the CDFM using results established in \citet{patel:2022}. We first need some notations and assumptions. 

We will assume that each mixing distribution $G_{k,\mvgl}$ for $k \in [K]$ is sub-Gaussian with the same sub-Gaussian parameter $\gs^2$ and write $G_{k,\mvgl}\sim subG(\gs^2)$. In other words, for $k \in [K]$ and $a \in \bR^r$,
\begin{equation}
\E_{G_{k,\mvgl}} e^{\langle \mvgl - \mvgm_k, a\rangle} \leq e^{\frac{\gs^2\|a\|_2^2}{2}}.
\end{equation}
Note that this assumption controls the concentration of $G_{k,\mvgl}$ even for $G_{k,\mvgl}$ supported on $\bD^{r-1}$. Let $\{\wh\mvgm\s_k\}$ and $\zhat\s$ be the estimated means and community membership function after $s$ iterations of Lloyd's algorithm applied to $\wh \mvgl_i$'s. Recall from Definition \ref{def:communities} that $z$ denotes the true community membership function.
Define the misclustering rate at step $s$ as 
\begin{equation}\label{eq:A_s} 
A_s = \dfrac{1}{d} \sum_{i=1}^d \cI \{ \zhat\s_i \neq z_i\}.
\end{equation}
Recall that $\mvgm_k$ denotes the mean of $G_{k,\mvgl}$ as defined is \eqref{def:mixture_means}. Define the minimal and maximal distance between the true cluster means as 
\begin{equation}
\gD = \min_{k\neq l} {\| \mvgm_k - \mvgm_l}\|_2 \quad \text{and} \quad M = \max_{k\neq l} {\| \mvgm_k - \mvgm_l}\|_2 .
\end{equation}
In view of \eqref{lambda max bound}, define 
\begin{equation}\label{def:eps}
\gee = \max_{i \in [d] } \|\wh\mvgl_i - \mvgl_i\|_{2},
\end{equation}
i.e.\ the maximal error between the true and estimated loadings. Let $\ga = \min_{k\in[K]}{\frac{n_k}{d}}$ be the minimum cluster size proportion and define the sub-Gaussian and estimation error signal-to-noise ratios as 
\begin{equation}\label{eq:signal-to-noise}
\gr_{\gs} = \dfrac{\gD}{\gs} \sqrt{\dfrac{\ga}{1 + \frac{Kr}{d}}} \text{ and } \gr_{\epsilon}=\frac{\sqrt{\ga}\gD}{\epsilon},
\end{equation}
respectively. 
Define $\delta(d,\sigma,\Delta,\epsilon)$ as
\begin{equation}\label{delta-pron}
    \delta(d,\sigma,\Delta,\epsilon) = \frac{1}{d}+2 \exp \left(-\frac{\Delta}{\sigma}\right)  + 2 \exp\left( -\frac{\Delta^2}{8\epsilon\sigma} \right)
\end{equation}
Note that $\gd \to 0$  as $\gr_{\gs},\gr_{\epsilon},d \to \infty$. 

Then, given that $G_{\mvgl}$ is a sub-Gaussian mixture, \citet{patel:2022}, building on the work of \citet{lu:2016}, prove Theorem \ref{our-lloyd-thm} below about the clustering accuracy of Lloyd's algorithm on $\wh\mvgL$ assuming good initialization. The condition on the initialization of Lloyd's algorithm requires either the cluster-wise misclustering rate or the maximal distance between the true and estimated cluster means to be small enough. 

More precisely, define $\wh z^{(0)}$ as the initial community membership function given to Lloyd's algorithm and set
\begin{equation}\label{eq:U set}
U^{(0)}_{lk} = \{i : z_i = l,\; \wh z^{(0)}_i = k\}
\end{equation}
to be the set of loadings in community $l \in [K]$ which are initially misclustered into community $k \neq l$. Define $\nhat^{(0)}_{lk} = |U^{(0)}_{lk}|$, $\nhat^{(0)}_k = \{i : \zhat^{(0)}_i = k\}$, and the cluster-wise misclustering as 
\begin{equation}\label{eq;G_s}
G^{(0)} = \max_{k \in [K]} \bigg\{ \dfrac{1}{\nhat^{(0)}_k} \sum_{l \neq k} \nhat^{(0)}_{lk},\; \dfrac{1}{n_k}\sum_{l\neq k} \nhat^{(0)}_{kl} \bigg\},
\end{equation}
where $n_k$ is the true size of community $k$. Then, an initial community membership $\wh z^{(0)}$ with initial means $\{\wh\mvgm^{(0)}_k\}_{k\in[K]}$ is considered good enough if either one of the following hold:
\begin{equation}
\begin{split}
G_0 &\leq \left(\dfrac{1}{2} - \frac{\sqrt{6}+1}{\rho_{\sigma}} - \frac{2.1\sqrt{\alpha}+1} {\rho_{\epsilon}} - \frac{1}{\alpha^{1/4}}\sqrt{\frac{\sigma}{\Delta}} \right)\dfrac{\Delta}{M}\\
\max_{k \in [K]} \dfrac{\|\wh\gm^{(0)}_k - \gm_k\|}{\gD} &\leq \frac{1}{2} - \frac{1}{\rho_{\sigma}} - \frac{1.1\sqrt{\alpha}+1} {\rho_{\epsilon}} - \frac{1}{\alpha^{1/4}}\sqrt{\frac{\sigma}{\Delta}}
\end{split}
\end{equation}
Note the initialization condition weakens as both signal-to-noise ratios increase.

\begin{thm}{[\citet{patel:2022}]}\label{our-lloyd-thm}
Assume that $d\alpha \geq C_{1} K\log d$, $\rho_{\sigma} \geq C_{2} \sqrt{K}$, $\rho_{\epsilon}  \geq C_3\sqrt{K}$, $\frac{\Delta^2}{\epsilon\sigma} \geq  r\log(3)$ for some sufficiently large constants $C_{1},C_{2},C_{3} >0$. Conditional on a good enough initialization, we have 
            \begin{equation}
             A_s \leq \max \left\{ \exp \left( -\dfrac{\Delta^2}{16 \sigma^2} \right) , \exp\left( -\frac{\Delta^2}{8\epsilon\sigma} \right) \right\} \quad \text{for all} \quad s\geq 4\log d
            \end{equation}
            with probability greater than $1 -\delta(d,\sigma,\Delta,\epsilon)$.
\end{thm}

Note that the bound on the misclustering rate depends on both the effective sub-Gaussian and estimation error signal-to-noise ratios. The final max term in the error bounds indicates that it is not good enough to have just one of the terms be small. We must have good estimation of the loadings and sub-Gaussian error relative to the distance between the means for Lloyd's algorithm to work well in this case. One may be puzzled by the disappearance of the length $T$ and the dimension $d$ of the time series, but note that $d$ and $T$ enter in $\gee$ through \eqref{lambda max bound}. Our simulations in Section \ref{s:simulated data} will examine the result of Theorem \ref{our-lloyd-thm} from the numerical standpoint.  

\subsubsection{Choice of $K$}\label{s:chooseK}
In practice, it is often the case that researchers choose $K$ to equal $r$ based on the scree plot of the eigenvalues of the correlation matrix. However, as shown in Example \ref{Uniform Circle Example}, it may be the case that $r \neq K$ and choosing $r=K$ may lose significant information. Thus, given $r$, we need suitable methodology for picking $K$. We use SigClust \cite{liu:2008}, a procedure which tests whether the data come from a single normal distribution to judge the significance of clustering. 

The procedure for SigClust is as follows. First, we initialize SigClust by assigning two communities to the data. We choose to do so with Lloyd's algorithm with $K=2$. Then SigClust simulates the data multiple times by sampling i.i.d.\ observations from the estimated null distribution. For each simulation, SigClust calculates the Cluster Index (CI), the sum of within community variation over the total variation, based on the initial community assignments. The simulated CI distribution is compared with the observed CI in the data. Lastly, a $p$-value obtained as a quantile from the empirical distribution of cluster indexes is given. 

We apply SigClust iteratively on estimated loadings. More precisely, we run SigClust on $\{\wh\mvgl_i\}_{i\in[d]}$, with initial two communities given by Lloyd's algorithm. If the outputted $p$-value is larger than some predetermined threshold $\gt$, we conclude and determine $K=1$. Otherwise, we split the estimated loadings into two groups based on the initialized community assignments. Then for each group we repeat the SigClust procedure separately. Once the procedure ends, we choose $K$ to be the number of groups the loadings were split into. Note that this procedure not only picks $K$ but also clusters the estimated loadings. However, we only use SigClust to choose $K$ in our work. In Section \ref{s:sigclust_sims}, we examine how SigClust performs for different values of $r,K,$ and $\gt$ in the CDFM. We are actively pursuing a principled approach to choosing both $K$ and $r$ simultaneously, but defer this to future work. 


\subsection{Parametric Estimation of Mixing Distributions}\label{s:PE}

We discuss below two parametric estimation methods when the mixing distributions are the restricted normal distributions as described in Section \ref{s:RND}: maximizing the log likelihood and noise contrastive estimation \cite{gutmann:2010}. In either case, we first perform community detection on $\wh\mvgL$ as in Section \ref{s:k-meansCDFM}. Then, for each community $k \in [K]$, we treat $\{\wh\mvgl_i : \wh z(i) = k\}$ as an i.i.d.\ sample from a restricted normal distribution. We then numerically optimize the log-likelihood or the objective function of noise contrastive estimation as described below.

\subsubsection{Maximizing Log-Likelihood}

Let $\mvY \sim \cN_r (\mvgm_Y, \mvgS_Y)$ with $\mvgS_Y $ positive definite and $\mvY_D$ be its restriction to $\bD^{r-1}$. Let $\mvY_1,\ldots,\mvY_n$ be the observed data generated as $\mvY_D$. The log-likelihood of the data is given by 
\begin{equation}\label{loglikelihood}
\sum_{i=1}^n \log f_{\mvY_D}(\mvY_i) = \sum_{i=1}^n ( \log f_{\mvY}(\mvY_i)  - \log \pr (\mvY \in \bD^{r-1}) ),
\end{equation}
where $f_{\mvY}$ is the pdf of $\mvY$ and $ \pr (\mvY \in \bD^{r-1}) = F_{\mvY\mvY'}(1)$ with $F_{\mvY\mvY'}$ given in \eqref{imhof-method}. One can maximize \eqref{loglikelihood} using numerical optimization techniques. 

\subsubsection{Noise Contrastive Estimation}\label{s:NCE}
Noise Contrastive Estimation (NCE) \cite{gutmann:2010} allows estimating the parameters of the normal distributions without the need to calculate $ \pr (\mvY \in \bD^{r-1})$. This is done by creating noise data similar to the observed data, $\mvY_1,\ldots,\mvY_n$, and building a model which can differentiate between the observed and noise data. Generate noise data $\mvV_1,\ldots,\mvV_n$ from a normal distribution with mean $\wb\mvY = \frac{1}{n}\sum_{i=1}^n \mvY_i$ and variance $\wh\mvgS_{\mvY} = \frac{1}{n-1}\sum_{i=1}^n (\mvY_i - \wb \mvY)(\mvY_i - \wb \mvY)'$. Let $f_{\mvY_D}$ and $f_{\mvV}$ denote the pdfs of $\mvY_D$ and the noise distribution $\cN(\wb\mvY,\wh\mvgS_Y)$, respectively. One treats $\log \pr(\mvY \in \bD^{r-1})$ as a constant $c$ throughout so that $f_{\mvY_D}$ is a parametric function of $\mvmu_Y$, $\mvgS_Y$, and $c$. One then maximizes (as a function of $\mvmu_Y$, $\mvgS_Y$, and $c$) 
\begin{equation}\label{nce_obj}
\dfrac{1}{2n} \sum_{i=1}^n \log \bigg (  \dfrac{f_{\mvY_D} (\mvY_i) }{f_{\mvY_D} (\mvY_i) + f_{\mvV}(\mvY_i)} \bigg ) + \log \bigg (  \dfrac{f_{\mvV} (\mvV_i) }{f_{\mvY_D} (\mvV_i) + f_{\mvV}(\mvV_i)} \bigg )
\end{equation}
with respect to $\mvmu_Y$, $\mvgS_Y$, and $c$. In Section \ref{s:NCE_sims}, we evaluate the performance of NCE for simulated data.


\section{Numerical Studies}\label{s:DS}

In this section, we assess community detection and other estimation methods for both simulated and real data. 

\subsection{Simulations}\label{s:simulated data}
We examine the result of Theorem \ref{our-lloyd-thm} for simulated data and show the effects of $d,T,$ and $\gr_{\gs}$ on misclustering. Also, we examine how well the SigClust procedure estimates $K$ when $r$ is known and examine performance of NCE in estimating  mixing means and variances.  

\subsubsection{Empirical Consistency of $k$-Means}\label{s:thm_sims}

We consider the case when $r=2$ for ease of visualization. We let the community sizes be equal so that $\ga = \frac{1}{K}$. In order to control sub-Gaussian signal-to-noise ratio $\gr_{\gs}$, we need to use distributions $G_{k,\mvgl}$ on $\bD^{r-1}$ whose sub-Gaussian parameter can be easily manipulated. We let $G_{k,\mvgl}$, $k = 1,\ldots,K$, be normal distributions restricted to $\bD^{r-1}$ centered at $\mvgm_k \in \bD^{r-1}$ with variance matrix given by $\gs_k^2 \mvI_r$ as detailed in Section \ref{s:RND}. Then $G_{k,\mvgl} \sim subG(\gs_k^2)$ by definition. Furthermore, we restrict $\{\mvgm_k\}$ to be equidistant by requiring that $\mvgm_k = m (\cos \frac{k2\pi}{K}, \sin \frac{k2\pi}{K})'$, for some $m \in (0,1)$, and we also suppose $\gs_k = \gs$ for all $k \in [K]$. Then $\gD = m\sqrt{3}$. So the sub-Gaussian signal-to-noise ratio in \eqref{eq:signal-to-noise} is given by 
\begin{equation}
\gr_{\gs} = \dfrac{m\sqrt{3}}{\gs^2} \sqrt{\dfrac{\ga}{1 + \frac{Kr}{d}}}.
\end{equation}

\begin{figure}[t]
\centering
\includegraphics[width=\textwidth,height = 1.9in]{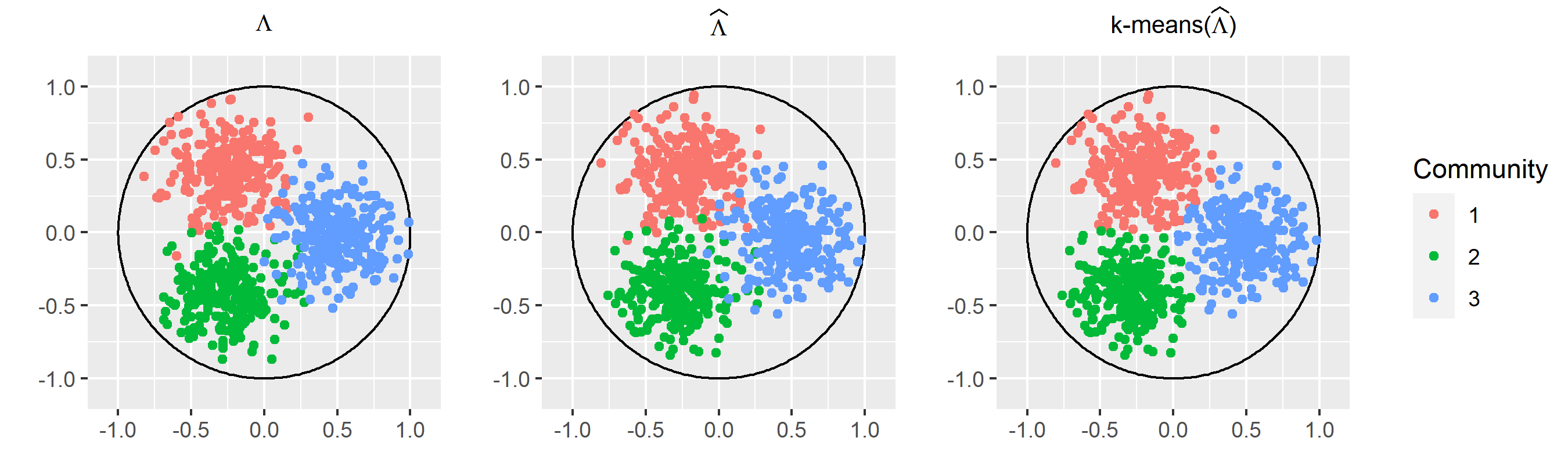}
\caption{Simulated loading vectors, their estimates, and the $k$-means community labels for the estimates when $(d,T,m,\gs) = (750,930,0.5,0.2)$. }\label{lambda-normal-example}
\end{figure}

Let $K=3$ and $S = \{ (d,T,m,\gs) : d = 30 + 90a, \; a =0,1,\ldots, 12, \;\;\; T= 30 + 90b, \; b =0,1,\ldots, 12, \;\;\; m = 0,.1,.2,\ldots,.9,  \;\;\; \gs = 0,.1,.2,\ldots,.9\}$ be a grid of parametric choices. For each $s \in S$, we construct loadings $\mvgL$ following the mixture distributions described above and generate i.i.d.\ factors $\{\mvf_t\}_{t=1,\ldots,T}$ from a multivariate normal distribution with mean $\mvzero$ and variance $\mvI_r$ (so there is no time dependence). The errors $\{\mvgee_t\}_{t=1,\ldots,T}$ are similarly sampled i.i.d.\ from a multivariate normal distribution with mean $\mvzero$ and variance $\mvI_d$.  We construct  $\{\mvX_t\}$ as in \eqref{DFM} and use the sample correlation matrix to obtain the PCA estimate $\wh\mvgL$ with $r = 2$. We then use Lloyd's algorithm to estimate the clustering $\wh z$. The misclustering rate is calculated given the true clustering. This procedure is repeated 10 times for each $s \in S$. An example of the procedure with $(d,T,m,\gs) = (750,930,0.5,0.2)$ is given in Figure \ref{lambda-normal-example}.

\begin{figure}[t]
\centering
\includegraphics[width = \textwidth]{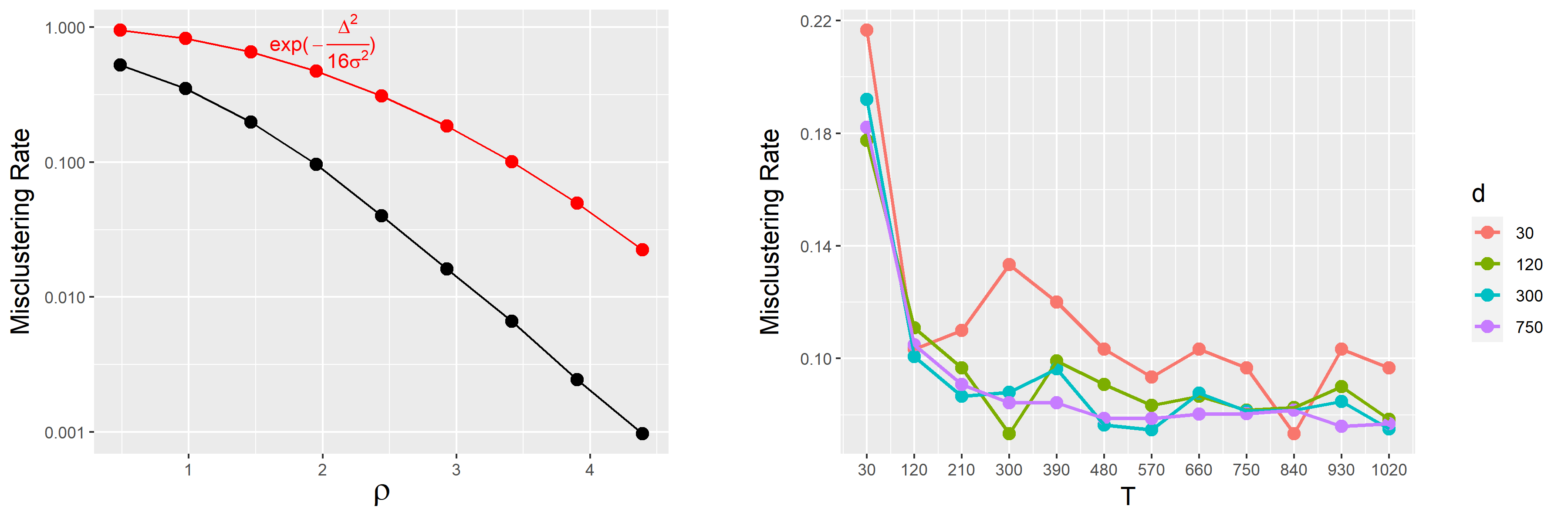}
\caption{Left: Average misclustering rates, on the log scale, for different values of $\gr_{\gs}$ given $\gs = 0.2$ and $d =120$. The red line indicates the value of $e^{-\frac{\gD^2}{16\gs^2}}$.   Right: Average misclustering rate of different values of $T$ and $d$.}\label{As_versus_Trho}
\end{figure}

\begin{figure}[t]
\centering
\includegraphics[width = \textwidth]{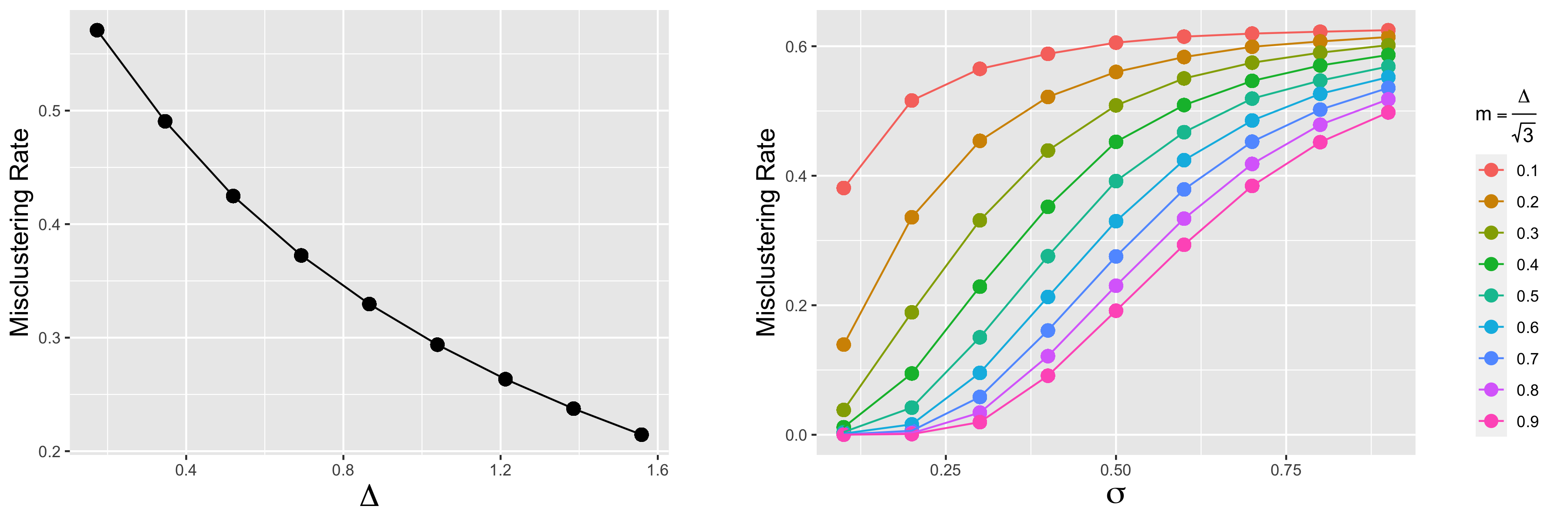}
\caption{Left: Average misclustering rates for different values of $\gD$.  Right: Average misclustering rates for different values of $\gs$ and $m$. }\label{As_versus_mgamma}
\end{figure}

The right plot of Figure \ref{As_versus_Trho} shows that the average misclustering rate decreases as $d$ or $T$ increases. Note that regardless of $T$, as $d$ increases, misclustering decreases. This is not surprising as larger samples from a sub-Gaussian concentrate around the true mean with high probability. The left plot of Figure \ref{As_versus_Trho} shows the average misclustering rate, on the log scale, for all simulations in which $\gs = 0.2$ and $d = 120$. As $\gr_{\gs}$ increases, the misclustering rate decreases and is below $e^{-\frac{\gD^2}{16\gs^2}}$, one of the terms in the bound obtained in Theorem \ref{our-lloyd-thm}. Taking other choices of $\gs$ and $d$ yields a similar plot. The individual effects of $m$ and $\gs$ on misclustering rate are shown in Figure \ref{As_versus_mgamma}. The right plot of Figure \ref{As_versus_mgamma} shows that as means get further apart and as noise gets smaller, misclustering rate decreases and the left plot shows that as the distance between means increases, misclustering rate decreases. 

\subsubsection{NCE}\label{s:NCE_sims}
\begin{figure}[t]
\centering
\includegraphics[width=\textwidth,height = 2in]{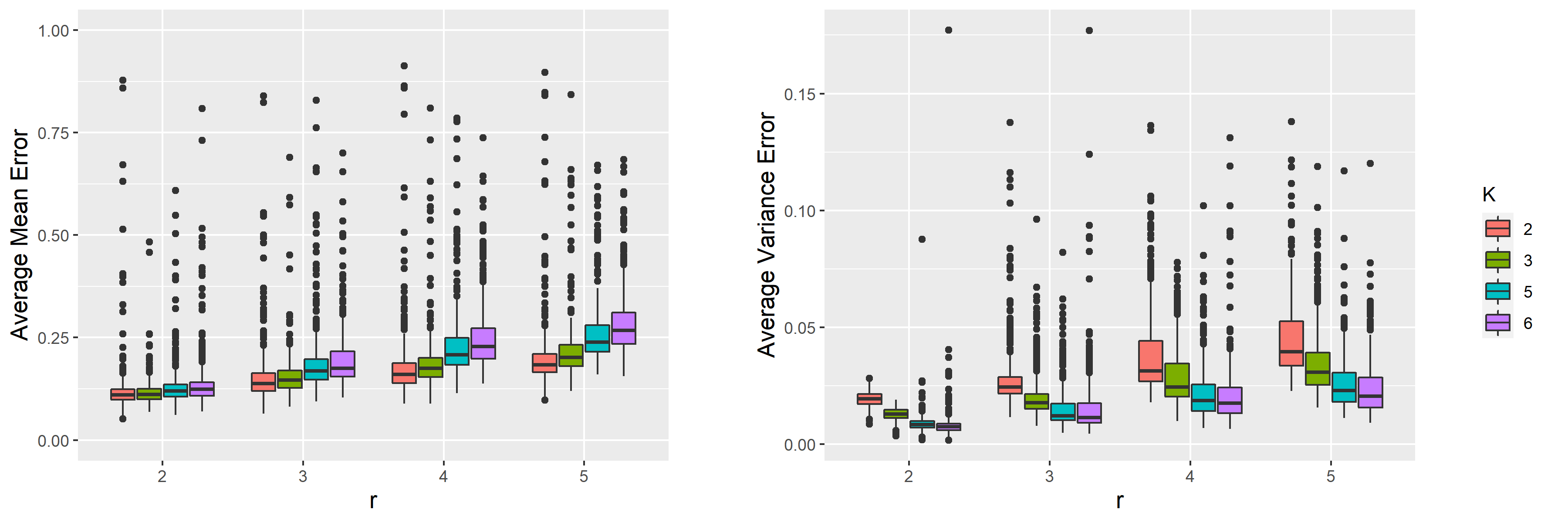}
\caption{Boxplots of average mean and variance estimation errors from NCE optimization across $r$ and $K$ with $(d,T,m,\gs) = (570,750,0.7,0.3)$. The true variance matrix for each mixing distribution is given by $\gs^2\mvI_{r} = .09\mvI_{r}$.}\label{f:NCE}
\end{figure}

We consider the same setup for the simulated data as in Section \ref{s:thm_sims} but with the following differences. We allow $r$ and $K$ to change but we fix $d = 570$ and $T = 750$. (These values of $d$ and $T$ were chosen to be comparable to \citet{gates:2016}.) We let $\mvgm_k = m \frac{\mvZ_k}{\|\mvZ_k\|} $, for some $m \in (0,1)$, where $\mvZ_k$ are i.i.d.\ and follow a $r$-dimensional standard normal distribution. The covariance matrix of the mixing distribution is given by $\gs^2 \mvI_r$. The estimation procedure for $\mvgL$ is the same as in Section \ref{s:thm_sims} with the exception that we assume the true community membership $z$ is known. For each cluster (community), we estimate the mean and variances (i.e. the diagonal of the covariance matrix) of the restricted normal distribution by numerically optimizing the NCE objective function \eqref{nce_obj} with initial $c = 1/2$. The mean error for each cluster is given by the $\ell_2$ norm of the estimated mean minus the true mean. The variance error for each cluster is given by the Frobenius norm of the estimated variances minus the true variances. The average mean and variance errors refer to average across clusters for each simulation. 

Figure \ref{f:NCE} shows the boxplots of the average mean error (left) and average variance error (right) for each choice of $r$ and $K$. Note that for both plots we have fixed $m = 0.7$ and $\gs = 0.3$. Thus, the true variance matrix is $\gs^2\mvI_r = .09\mvI_r$. Each boxplot is obtained from 400 simulations for that value of $r$ and $K$. Note that as $r$ and $K$ increase, the average mean error increases and variation across simulations grows. However, the average variance error decreases as $K$ increases. 

\subsubsection{SigClust}\label{s:sigclust_sims}

\begin{figure}[t]
\centering
\includegraphics[width=\textwidth,height = 4in]{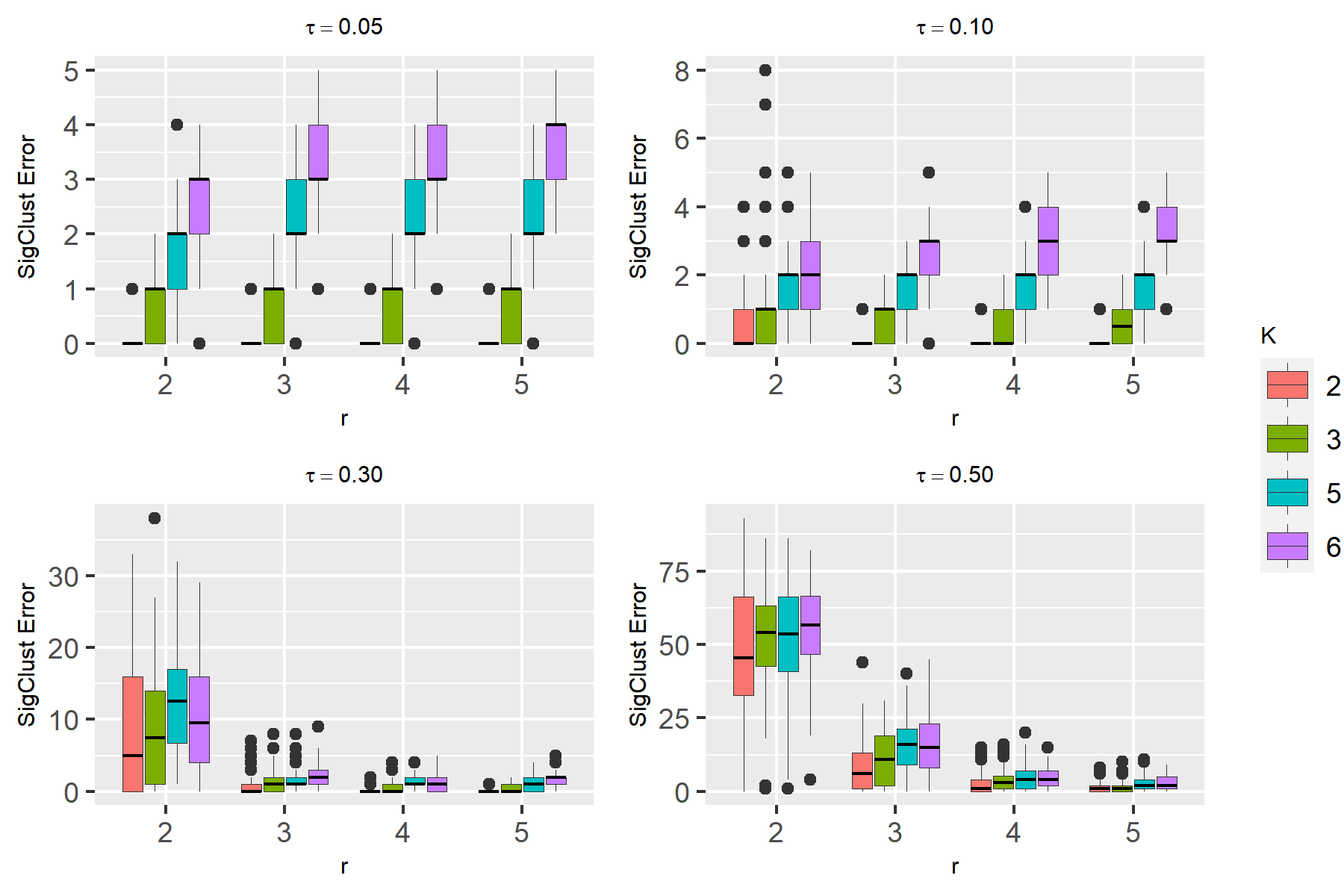}
\caption{Boxplots of SigClust error defined as $| \wh K_{\gt} - K |$ for different values of the $p$-value threshold $\gt$, number of factors $r$ and number of communities $K$ with $(d,T,m,\gs) = (750,930,0.7,0.3)$. }\label{f:SigClust}
\end{figure}

The simulations in Section \ref{s:thm_sims} and \ref{s:NCE_sims} assume that the true $K$ is known. In this section, we estimate $K$ using the SigClust procedure described in Section \ref{s:chooseK}. We consider the same setup as that in Section \ref{s:NCE_sims} and we let $\tau$ be the $p$-value threshold for the SigClust procedure. We define the SigClust error as $| \wh K_{\gt} - K |$ where $\wh K_{\gt}$ is the output of the SigClust procedure with threshold $\gt$. 

Figure \ref{f:SigClust} shows the boxplots of the SigClust error for different choices of $r$, $K$, and $\gt$ with $m = 0.7$ and $\gs = 0.3$. For each choice of $r$, $K$, and $\gt$, boxplots of 100 iterations of the simulation are plotted. Naturally, as $K$ increases the SigClust error also increases for all values of $\gt$. Note that as $r$ grows, larger values of $\gt$ do as well if not better than small $\gt$ seeming to indicate SigClust is conservative in these settings. Further explorations are deferred to future work.

\subsection{Applications}

\subsubsection{fMRI data}

We explore our $k$-means clustering procedure on fMRI data obtained from a subset of the Human Connectome Project \cite{elam:2021}. The fMRI time series are the BOLD (blood oxygen level dependent) signals for regions of interest (ROIs) in a brain. The data here include fMRI time series for 10 individuals who are going through multiple tasks with short resting states between tasks. There are 268 ROIs with 392 time points corresponding to $784$ seconds. We focus on one subject's fMRI scan and restrict ourselves to 58 ROIs. So, our time series has dimension $d = 58$ and length $T = 392$. 

\begin{figure}
\centering
\includegraphics[width = .9\textwidth,height = 1.8in]{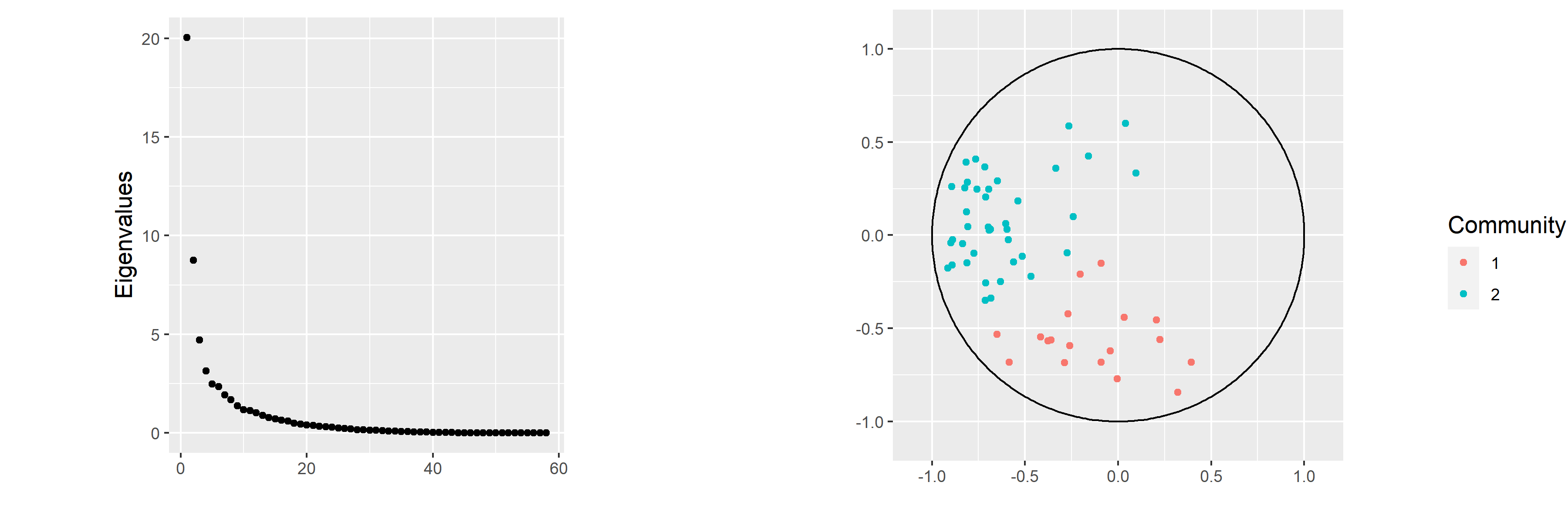}
	\caption{Left: Eigenvalues of the sample correlation matrix of the fMRI data.  Right:  Loading vectors in the first two PCA dimensions colored by $k$-means community labels.}\label{fMRI-sample-corr}
\centering
\includegraphics[width = \textwidth]{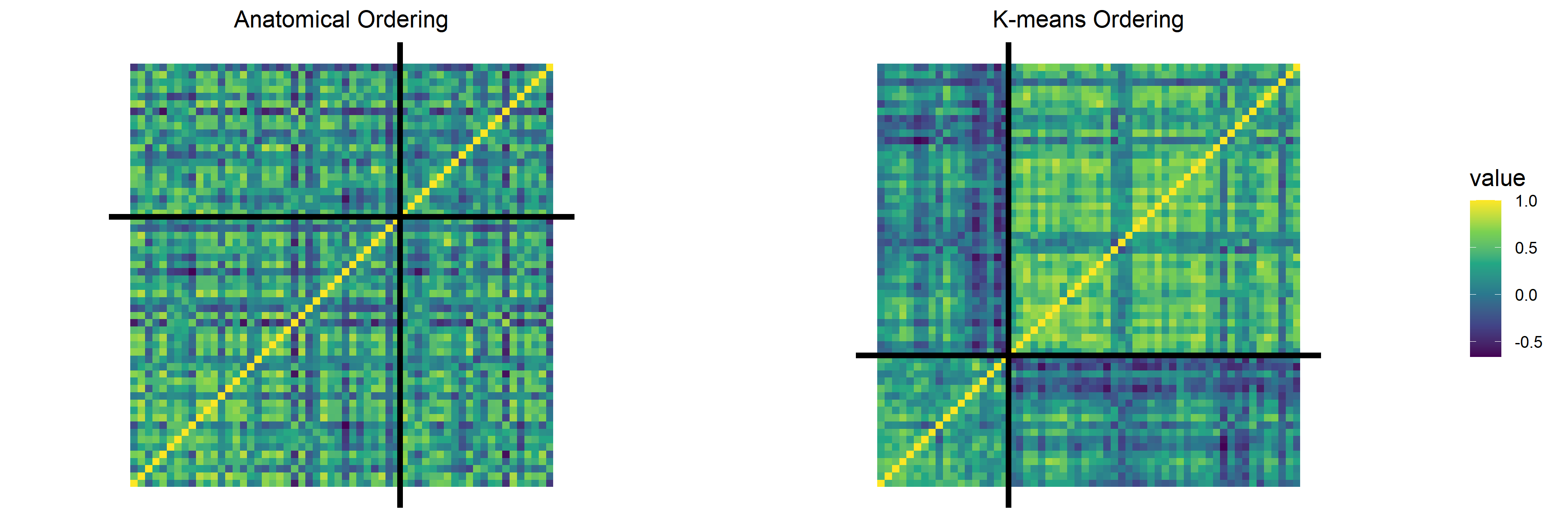} 
   \caption{Left: Sample correlation matrix reordered using the Default Mode Network (bottom left block) and Attention Network (top right block). Right: Sample correlation matrix after reordering of the nodes using $k$-means community labels with $K = 2$ chosen by the SigClust procedure with a threshold of $\gt = 0.30$. }\label{fMRI-loadings-corr}
\centering
\includegraphics[width = \textwidth]{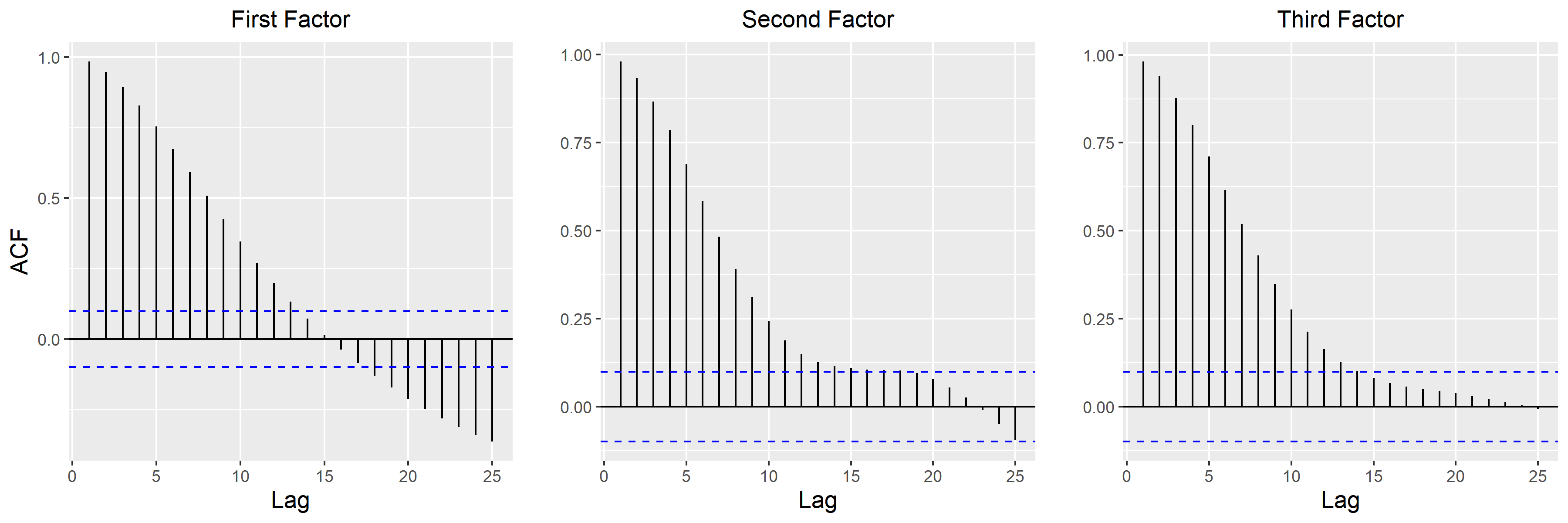}
	\caption{ACFs for each of the three factors estimates for the fMRI data. }\label{lambda-normal-ACF}
\end{figure}

 The eigenvalues of the sample correlation matrix and the PCA estimates $\wh\mvgL$, with $r = 3$ and community labels given by Lloyd's algorithm with $K=2$, are plotted in Figure \ref{fMRI-sample-corr}. Figure \ref{fMRI-loadings-corr} shows the sample correlation matrix with two different orderings. The black lines indicate separation by community. The first ordering is with respect to the Default Mode Network (DMN) and Attention Network (AN) which are two different brain networks consisting of several structural regions of the brain. The DMN is known to be one of the regions of the brain which are deactivated when during directed tasks. The AN, on the other hand, is known to be active during directed tasks. The second ordering uses the community labels obtained from Lloyd's algorithm with $K = 2$, which was the estimated number of clusters using the SigClust procedure with $r = 3$ and $\gt = 0.30$. Note that there is a clear block structure once we reorder using our estimated communities. In general, fMRI time series have temporal dependence. Figure \ref{lambda-normal-ACF} includes the sample auto-correlation functions (ACFs) of the $3$ estimated factor series.

\subsubsection{Macroeconomic Data}

We perform community detection on the US Quarterly data taken from the DRI/McGraw-Hill Basic Economics database of 1999\footnote{\url{https://dataverse.unc.edu/dataset.xhtml?persistentId=hdl:1902.29/D-17267}} which consists of $d = 144$ macroeconomic time series sampled quarterly from 1959 to the end of 2006 resulting in $T = 190$. A more detailed description of the time series along with the transformations to stationarity that have been made are provided in the Data Appendix of \citet{stock:2009}. 

\begin{figure}[t]
\centering
\includegraphics[width = \textwidth]{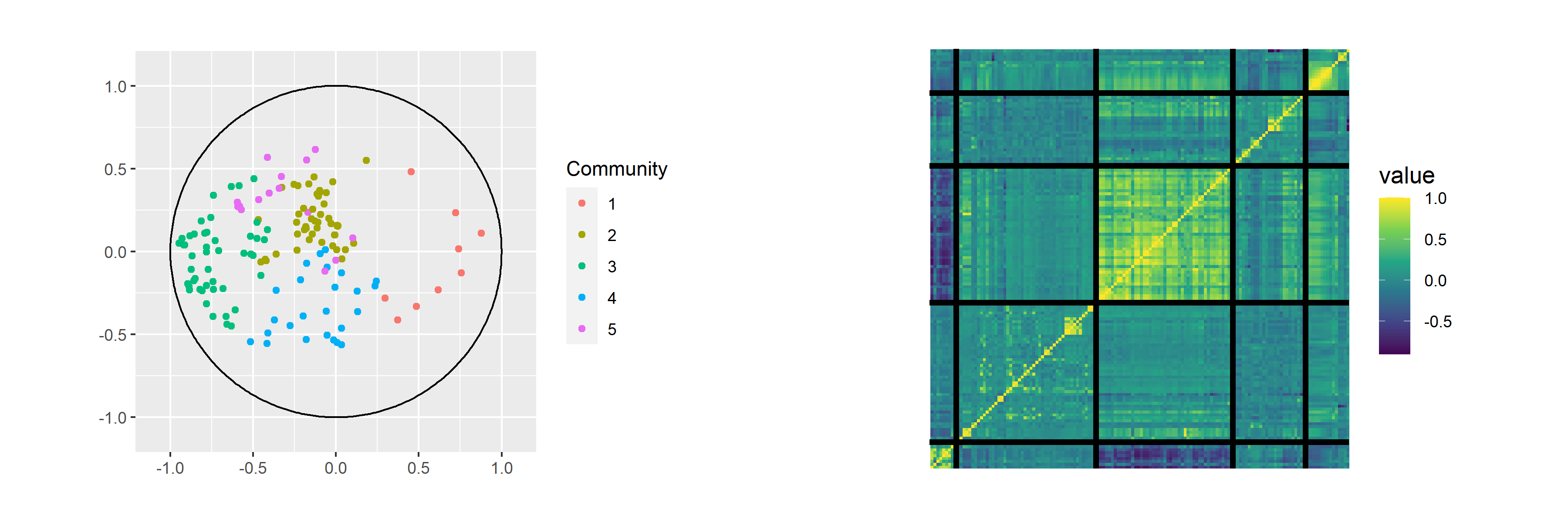} 
   \caption{Left: Projected estimated loadings $\wh\mvgL$ on to the first two dimensions.  Right: Sample correlation matrix reordered using the $k$-means community labels.  }\label{USquarterly figure}
\centering
\includegraphics[width = \textwidth]{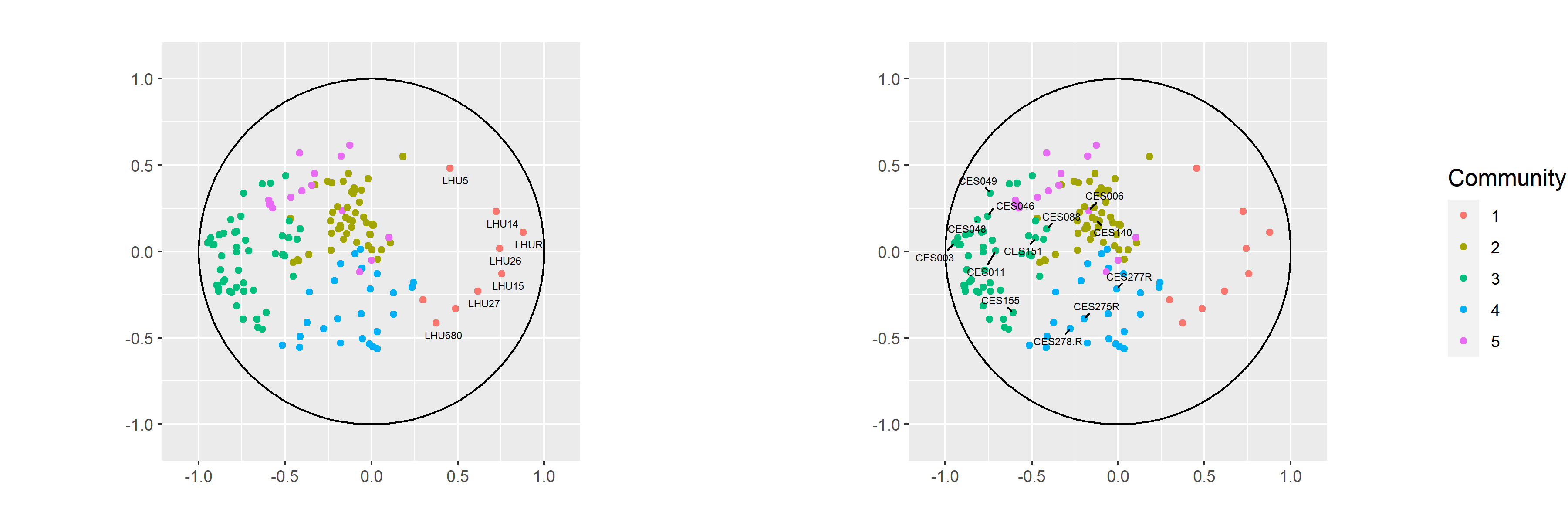} 
   \caption{ Projected estimated loadings $\wh\mvgL$ on to the first two dimensions with labels for ``LHU" and ``CES" labeled time series on the left and right, respectively. }\label{USquarterly labelled figure}
\end{figure}

We estimate the loading matrix $\wh\mvgL$ using $r = 4$ factors and perform $k$-means with $K = 5$ communities. The number of communities was chosen by the SigClust procedure with $\gt = 0.30$. The loadings projected to the first two dimensions are plotted in Figure \ref{USquarterly figure} where the communities are distinguished using color. Figure \ref{USquarterly figure} also shows the sample correlation matrix whose entries are reordered according to the $k$-means algorithm. The sample correlation matrix exhibits block structure and the projection of the $\wh\mvgl_i$'s shows separation. Some of the more interesting clusters are shown in Figure \ref{USquarterly labelled figure} with labels for the time series. All the time series with labels starting with ``LHU" deal with unemployment rate. The time series starting with ``CES" with a number less than 140 represent number of employees in different categories. The time series labeled ``CES151" represents average weakly hours and overtime hours, respectively. The rest of the ``CES" time series end with ``R" and represent real average hourly earning for different populations and we can see a clear separation between the ``CES" series with and without ``R" in the loadings in Figure \ref{USquarterly labelled figure}.

\section{Connections of CDFMs to Other Constructions}\label{s:COC}

\subsection{Random Dot Product Graphs}\label{s:RDPG}
Latent position random graphs \cite{hoff:2002} allow for random heterogeneous node attributes which determine relationships (edge connections) between nodes. More specifically, each node $i$ has attribute $a_i$ sampled from some latent space $\cX$ and the probability of an edge between nodes $i$ and $j$ is given by $\gk(a_i,a_j)$ for some kernel function $\gk$. One concrete example in this general class is as follows: take $\cX$ to be any subset of $\bR^r$ such that for all $\mvx,\mvy \in \cX$, $\mvx'\mvy \in [0,1] $ and $\gk$ to be the dot product function. The corresponding latent position random graph is called the Random Dot Product Graph (RDPG) \cite{young:2007}. Denote $\mathbfcal{A} \in \bR^{d\times r}$ as the matrix of attributes for $d$ nodes with $i$-th row given by $\mva_i'$.  The matrix of probabilities between edges is given by $\mvP = (P_{i,j}) = \mathbfcal{A}\mathbfcal{A}'$. The adjacency matrix $\mvA = (A_{i,j})$ of the RDPG is given by $A_{i,j} \sim$ $Bernoulli(P_{i,j}$).

Viewing the matrix of attributes $\mathbfcal{A}$ as the loading matrix $\mvgL$, one can find a number of similarities between CDFMs and RDPGs. The average connection probability matrix is $\mvP = \mathbfcal{A}\mathbfcal{A}'$ for the RDPG, while the average weight matrix is $\mvgL\mvgL'$ for the CDFM. Furthermore, the adjacency matrix $\mvA$ is a noisy version of $\mvP$, as $\wh\mvgS_X$ is for $\mvgS_X$. RDPGs have the same identifiability issue, as for any invertible $\mvC$, $\mvP = \mathbfcal{A}\mvC\mvC^{-1}\mathbfcal{A}'$. The RDPG can also exhibit a block structure by controlling how the attributes are sampled from $\cX$.

Methods used to estimate the latent attributes in the RDPG literature use spectral embedding of the adjacency matrix, similar to our PCA estimate of the loading matrix. 
Community detection for RDPGs is often done through methods created for Stochastic Block Models (SBMs), a special case of RDPGs where nodes within the same community have the same latent attributes. \citet{rohe:2011} show that spectral clustering on the degree-corrected adjacency matrix offers consistent community detection for SBMs. Under some eigenvalue conditions and enough separability in the attributes among different communities, \citet{lyzinski:2014} show that $k$-means clustering on the spectral embedding of $\mvA$ can achieve perfect clustering with high probability. See the survey paper by \citet{athreya:2018} for an overview of such questions. 

Despite the similarities, there are also differences from the RDPG literature. The obvious differences in the range of attributes (e.g.\ those for CDFMs can be negative) and the observed values ($0$-$1$ versus real values) aside, when considering communities in RDPGs, attributes have thus far been assumed to be constant (i.e.\ taking the same value) within communities (see \citet{athreya:2018} and references therein). We do not make such assumption for CDFMs, allowing the attributes to be drawn from non-degenerate mixing distributions $G_{k,\mvgl}$. This flexibility might be more important for CDFMs than for RDPGs, since even constant attributes for the latter model result in significant variability of the adjacency matrix. 

\subsection{Reduced-Rank and Network VAR Models}

The CDFM model is also closely related to VAR models with network structure. For these models, the network structure is imposed through the autoregressive coefficients, and the dynamics of the series are driven by network properties such as community structure, similarly to the CDFM. One such VAR model studied by \citet{gudmundsson:2017} is defined as follows. Focusing on the VAR$(p)$ model of order $p=1$, and symmetric transition matrix for simplicity, consider
\begin{equation}\label{WSBM-VAR}
\mvY_t = \mvPsi \mvY_{t-1} + \mvxi_t,
\end{equation}
where $\E\mvxi_t = \mv0$, $\E\mvxi_t\mvxi_s = \mv0$ for $t\neq s$ and $\E\mvxi_t\mvxi_t' = \mvI_d$, i.e.\ $\{\mvxi_t\}\sim$ WN$(\mv0,\mvI_d)$. Furthermore,
\begin{equation}\label{WSBM-VAR-coeff}
\mvPsi = \varphi \mvD^{-1/2}\mvA\mvD^{-1/2},
\end{equation}
with $\varphi > 0 $, symmetric $\mvA \in \bR^{d\times d}$ and diagonal $\mvD \in \bR^{d\times d}$. The matrix $\mvA = (A_{i,j})_{i,j\in [d]}$ is viewed as an adjacency matrix of a weighted, undirected network with nodes $i,j \in V := [d]$:
\begin{equation}
A_{i,j} = A_{j,i} =
    \begin{cases}
        0 & \text{if no edge b/w \text{nodes} $i$ and $j$,} \\
        w_{i,j} & \text{if an edge b/w \text{nodes} $i$ and $j$,}
    \end{cases}
\end{equation}
where $w_{i,j}$ are weights. The diagonal elements $D_{i,i}$ of $\mvD$ are the node degrees $D_{i,i} = \sum_{j\neq i} A_{i,j}$. To have a community structure with $K$ communities, let $\mvZ = (Z_{i,k})$ be the $d\times K$ community membership matrix with $Z_{i,k} = 1$ if node $i$ belongs to community $k$ and let $z: [d] \to [K]$ be the community assignment function such that $z(i) = k$ if $Z_{i,k} =1$ and $= 0$ otherwise. Let $\mvB\in \bR^{K\times K}$ be a symmetric matrix with edge probabilities among communities. In a weighted stochastic block model (WBSM), the probability of an edge between nodes $i$ and $j$ is taken as $B_{z(i),z(j)}$. A random weight $w_{i,j}$ is then drawn from a distribution $F_W$ on an interval $[\ga,\gb]$, $\ga,\gb \in \bR$.\footnote{More generally, \citet{gudmundsson:2017} consider VAR models of arbitrary order $p$ and also allow for degree correction in an SBM.}
We will refer to the VAR$(1)$ model \eqref{WSBM-VAR}--\eqref{WSBM-VAR-coeff} as the WSBM-VAR.

After taking the average, or expected value, with respect to the WSBM network, the WSBM-VAR becomes a reduced-rank VAR as follows. We have $\E \mvA = \gm_W \mvZ\mvB\mvZ'$, with $\gm_W = \E w_{i,j}$, and $\E \mvD = d\wb\mvD$ where $\wb\mvD = diag(C_{z(i)}:i\in[d])$ with $C_k$, $k\in [K]$. Replacing $\mvA$ and $\mvD$ by their expected values in \eqref{WSBM-VAR-coeff}, the transition matrix $\mvPsi$ can be thought as 
\begin{equation}\label{SBM-VAR-coeff}
 \wb\mvPsi = \dfrac{\varphi\gm_W}{d} \wb\mvD^{-1/2}\mvZ\mvB\mvZ'\wb\mvD^{-1/2},
\end{equation}  
and the corresponding VAR$(1)$ process as 
\begin{equation}\label{SBM-VAR}
\wb\mvY_t = \wb\mvPsi \wb\mvY_{t-1} + \mvxi_t.
\end{equation}
Note that the transition matrix $\wb\mvPsi$ is of reduced rank $K$. 

The VAR model \eqref{SBM-VAR} can be rewritten as a CDFM as follows. Define 
\begin{equation}
\mvgL = \dfrac{\varphi\gm_W}{d^{1/2}}\wb\mvD^{-1/2} \mvZ \quad \text{and} \quad \mvf_t = \dfrac{1}{d^{1/2}}\mvB\mvZ'\wb\mvD^{-1/2}\wb\mvY_{t-1}.
\end{equation}
Then,
\begin{align}
\wb\mvY_t &= \wb\mvPsi \wb\mvY_{t-1} + \mvxi_t \nonumber \\
& =   \dfrac{\varphi\gm_W}{d} \wb\mvD^{-1/2}\mvZ\mvB\mvZ'\wb\mvD^{-1/2}\wb\mvY_{t-1} + \mvxi_t \nonumber \\
& = \mvgL\mvf_t + \mvxi_t, \label{associatedDFM}
\end{align}
with 
\begin{align}
\mvf_t &= \dfrac{1}{d^{1/2}}\mvB\mvZ'\wb\mvD^{-1/2}\wb\mvY_{t-1} \nonumber \\
& =  \dfrac{1}{d^{1/2}}\mvB\mvZ'\wb\mvD^{-1/2}(\wb\mvPsi \wb\mvY_{t-2} + \mvxi_{t-1}) \nonumber \\
& =  \dfrac{1}{d^{1/2}}\mvB\mvZ'\wb\mvD^{-1/2}\dfrac{\varphi\gm_W}{d} \wb\mvD^{-1/2}\mvZ\mvB\mvZ'\wb\mvD^{-1/2} \wb\mvY_{t-2} + \dfrac{1}{d^{1/2}}\mvB\mvZ'\wb\mvD^{-1/2}\mvxi_{t-1} \nonumber \\
& =  \dfrac{1}{d}\mvB\mvZ'\wb\mvD^{-1}\mvZ\mvf_{t-1} + \dfrac{1}{d^{1/2}}\mvB\mvZ'\wb\mvD^{-1/2}\mvxi_{t-1} \nonumber \\
& = \mvgF \mvf_{t-1} + \mvgh_t, \label{associatedfactors}
\end{align}
where $\mvgF :=  \frac{1}{d}\mvB\mvZ'\wb\mvD^{-1}\mvZ$ and $\{\mvgh_t :=\frac{1}{d^{1/2}}\mvB\mvZ'\wb\mvD^{-1/2}\mvxi_{t-1} \} \sim$WN$(\mv0,\frac{1}{d}\mvB \mvZ'\wb\mvD^{-1}\mvZ \mvB)$. Thus, the VAR model \eqref{SBM-VAR} can be viewed as a CDFM with a VAR$(1)$ factor series. The loading matrix $\mvgL$ is such that each row is drawn from a mixture distribution on points $\frac{\varphi\gm_W}{d^{1/2}} (0,\ldots,C_{z(i)}^{-1/2},\ldots,0) \in \bR^K$. Note that for $i,j \in [d]$, if $z(i) = z(j)$, then $C_{z(i)}= C_{z(j)}$. So each row of $\mvgL$ is drawn from a $K$-mixture. 

Although the reduced-rank VAR model \eqref{SBM-VAR} was rewritten as a CDFM in \eqref{associatedDFM}--\eqref{associatedfactors}, the resulting model is in fact unlike the CDFMs considered previously. Note that $\mvgL'\mvgL = \frac{\varphi^2\gm_W^2}{d} \mvZ'\wb\mvD^{-1}\mvZ$ is a $K\times K$ diagonal matrix with entries given by $\{\frac{n_k \varphi^2\gm_W^2}{d} C_{k}^{-1} : k\in[K]\}$, where $n_k$ is the size of community $k$. Using the notation of \eqref{factors-strong} and \eqref{factors-weak}, we therefore expect in this case that
\begin{equation}\label{factors-weak-very}
	\mvgL'\mvgL \asymp 1,
\end{equation}
that is, the CDFM \eqref{associatedDFM}--\eqref{associatedfactors} is associated with {\it very weak factors}, the case not covered by Proposition \ref{weak-factor-thm} which assumes \eqref{factors-strong} or \eqref{factors-weak}. Such connections between DFMs and reduced-rank VARs are studied in more detail in \citet{bhamidi:2022}. 
 

\section{Conclusion}

We introduced a community version of a dynamic factor model, by assuming that rows of a loading matrix are drawn from a mixture distribution. A classical community detection algorithm based on $k$-means applied to the sample correlation matrix, viewed as a weighted network, was shown to recover the communities with a specified misclustering rate. The model and community detection were examined on simulated and real data. 

Several questions related to the model remain open. For example, we are currently exploring change point detection methods assuming community structure is allowed to change with time. Investigating other methods than PCA for loadings estimation, e.g.\ sparse estimation, is another interesting direction to pursue. 


\appendix
\section{Assumptions for DFM} \label{Appendix-assumps}

For the sake of completeness, we include here the assumptions behind Proposition \ref{weak-factor-thm} on DFMs taken from \citet{uematsu:2019}. We first establish some necessary definitions and notation. Let $\mvF = (\mvf_1,\ldots,\mvf_T)'$ be the factor matrix and $\mvE = (\mvgee_1,\ldots ,\mvgee_T)'$ be the error matrix. 
Define $\gt = \frac{\log T}{\log d}$ so that $T = d^{\gt}$. The constant $\nu$ below is a fixed large constant.

\begin{assumption}[\textbf{Latent factors}]
The factor matrix $\mvF$ is specified as the vector linear process $\mvf_t = \sum_{\ell=0}^\infty \mvgP_\ell \mvgz_{t-\ell}$, where $\mvgz_t \in \bR^r$  are vectors of i.i.d.\ $subG(\gs_{\gz}^2)$ entries with standardized second moments and $\sum_{\ell=0}^\infty \mvgP_\ell \mvgP_\ell' = \mvI_r$. Moreover, there are $C_f>0$ and $\ell_f\in\bN$ such that $\|\mvgP_\ell\|_2\leq C_f \ell^{-(\nu+2)}$ for all $\ell\geq \ell_f$. 
\end{assumption}

\begin{assumption}[\textbf{Factor loadings}] \label{3-LFest-assump-sparseloadings}
$\mvgL$ is possibly sparse in the sense that there exists $\ga_j$, $j\in [r]$, such that the number of nonzero elements in $j$-th column is given by $d^{\ga_j}$ for $j \in [r]$ and $0 < \ga_r \leq \ldots \leq \ga_1 \leq 1$. Furthermore, $\mvgL'\mvgL $ is a diagonal matrix with entries $\gd^2_jd^{\ga_j}$, $j \in [r]$, with $0 < \gd_r d^{\ga_r/2} \leq \ldots \leq \gd_1 d^{\ga_1/2}< \infty$ such that if $\ga_j = \ga_{j-1}$ for some $j$, then there exists some constant $g >0$ such that $ \gd^2_{j-1} - \gd^2_j \geq g^{\frac{1}{2}} \gd^2_{j-1}$. 
\end{assumption}

\begin{assumption}[\textbf{Idiosyncratic errors}]
The error matrix $\mvE$ is independent of $\mvF$ and is specified as the vector linear process $\mvgee_t = \sum_{\ell=0}^\infty \mvPsi_\ell \mvxi_{t-\ell}$, where  $\mvxi_t \in \bR^d$ are vectors of i.i.d.\ $subG(\gs_{\xi}^2)$ entries, and $\mvPsi_0$ is a nonsingular, lower triangular matrix. Moreover, there are $C_e>0$ and $\ell_e\in\bN$ such that $\|\mvPsi_\ell\|_2\leq C_e \ell^{-(\nu+2)}$ for all $\ell\geq \ell_e$. 
\end{assumption}

In the proof of Proposition \ref{weak-factor-thm}, \citet{uematsu:2019} also assume that $\ga_r + \tau > 1$ and $\ga_1 + \frac{ \max \{1, \tau\}}{2} < \ga_r + (\ga_r \wedge \tau)$. 
Assumption 2 allows the loadings $\mvgL$ to be sparse and ensures that $\mvgL'\mvgL$ is a diagonal matrix whose entries satisfy a gap condition. Note that when $\ga_1 = \cdots = \ga_r = 1$ in Assumption 2, we are in the strong factor model setting expressed through \eqref{factors-strong}. Otherwise, the DFM is a weak factor series expressed through \eqref{factors-weak}. In this sense, Assumption 2 is a weaker assumption than typically made.

\bibliographystyle{plain}

\flushleft
\begin{tabular}{l}
Shankar Bhamidi, Dhruv Patel, Vladas Pipiras, Guorong Wu \\
Dept.\ of Statistics and Operations Research  \\
UNC at Chapel Hill \\
CB\#3260, Hanes Hall \\
Chapel Hill, NC 27599, USA \\
{\it bhamidi@email.unc.edu, dhruvpat@live.unc.edu, pipiras@email.unc.edu, guorong\_wu@med.unc.edu} \\
\end{tabular}


\end{document}